\documentclass[10pt] {article}

\usepackage[usenames,dvipsnames,svgnames,x11names]{xcolor}
\usepackage{color}
\definecolor{gray}{rgb}{0.4,0.4,0.4}
\definecolor{darkblue}{rgb}{0.0,0.0,0.6}
\definecolor{maroon}{rgb}{0.5,0,0}
\definecolor{darkgreen}{rgb}{0,0.6,0}
\usepackage{bbding}
\usepackage{latexsym,amsthm,amsfonts,amsmath,mathrsfs}

\DeclareMathOperator*{\subto}{s.t.}
\DeclareMathOperator*{\sel}{select}

\newcommand{\revnote}[1]{ {\textcolor{teal} { ***Reviewer: #1 }}} % needs a response

\newcommand{\bluecol}[1]{\textcolor{darkblue}{#1}}

\newcommand{\claim}[1]{\vspace{-0.08in}\subsubsection*{\textnormal{\textit{\bluecol{#1}}}}}

\newif\ifsectionbreaks
\sectionbreaksfalse

\usepackage[T1]{fontenc} 
\usepackage[normalem]{ulem}

\usepackage[export]{adjustbox}

\newcommand{\modc}[1]{{\textcolor{blue}{#1}}}
\newcommand{\addc}[1]{{\textcolor{darkgreen}{#1}}}
\newcommand{\delc}[1]{ {\textcolor{gray} {\sout{#1}} }}
\newcommand{\repc}[2]{ {\textcolor{gray} {\sout{#1}} }{\textcolor{darkgreen} {#2}}}
\newcommand{\tempdel}[1]{}

\renewcommand{\modc}[1]{#1}
\renewcommand{\addc}[1]{#1}
\renewcommand{\delc}[1]{}
\renewcommand{\repc}[2]{#2}

\renewcommand{\revnote}[1]{} 

\usepackage{balance}

\usepackage{listings}

\lstdefinelanguage{XML}
{
basicstyle=\ttfamily\footnotesize,
  morestring=[b]",
  moredelim=[s][\bfseries\color{Maroon}]{<}{\ },
  moredelim=[s][\bfseries\color{Maroon}]{</}{>},
  moredelim=[l][\bfseries\color{Maroon}]{/>},
  moredelim=[l][\bfseries\color{Maroon}]{>},
  morecomment=[s]{<?}{?>},
  morecomment=[s]{<!--}{-->},
  commentstyle=\color{gray},
  stringstyle=\color{blue},
  identifierstyle=\color{red}
}

\usepackage{moreverb}

\usepackage[nounderscore]{syntax}

\usepackage{wrapfig}

\graphicspath{{./figures/}}

\DeclareGraphicsExtensions{.pdf}
\usepackage[export]{adjustbox}

\usepackage[
 font=footnotesize
 ]{subfig} %[caption=false,font=footnotesize,labelfont=sf,textfont=sf]

\usepackage{algorithmicx}
\usepackage{algpseudocode}
\usepackage[ruled]{algorithm}
\usepackage{setspace}
\algrenewcommand{\algorithmiccomment}[1]{\hskip3em{{\footnotesize \textcolor{light-gray}{$\blacktriangleright$}}} #1}

\usepackage{multirow} % Multi-row tables
\usepackage{rotating} % sideways table
\usepackage{booktabs} % better lines
\usepackage{colortbl} % cell colors
\usepackage{tablefootnote} % add support for footnote in table
\usepackage{array} % table col alignment with width
\newcolumntype{L}[1]{>{\raggedright\let\newline\\\arraybackslash\hspace{0pt}}m{#1}}
\newcolumntype{C}[1]{>{\centering\let\newline\\\arraybackslash\hspace{0pt}}m{#1}}
\newcolumntype{R}[1]{>{\raggedleft\let\newline\\\arraybackslash\hspace{0pt}}m{#1}}

\usepackage[pdftex,colorlinks=true,urlcolor=blue,citecolor=blue]{hyperref}

\usepackage{xspace}

\usepackage{enumitem}

\usepackage{blindtext}

\usepackage{stfloats}

\newcommand{\mobilenet}{MobileNet\xspace}
\newcommand{\resnet}{ResNet\xspace}
\newcommand{\yolo}{YOLO\xspace}
\newcommand{\lstm}{LSTM\xspace}
\newcommand{\bert}{BERT\xspace}

\begin{document}

\title{PowerTrain: Fast, Generalizable Time and Power Prediction Models to
Optimize DNN Training on Accelerated Edges\thanks{~Preprint of article in Future Generation Computer Systems (FGCS): Prashanthi S. K., Saisamarth Taluri, Beautlin S, Lakshya Karwa, and Yogesh Simmhan, “PowerTrain: Fast, Generalizable Time and Power Prediction Models to
Optimize DNN Training on Accelerated Edges,” in \textit{Future Generation Computer Systems}, Elsevier, 2024, \href{https://doi.org/10.1016/j.future.2024.07.001}{doi:10.1016/j.future.2024.07.001}}}

\author{Prashanthi S. K., Saisamarth Taluri, Beautlin S,\\
Lakshya Karwa and Yogesh Simmhan\\
Department of Computational and Data Sciences (CDS),\\
Indian Institute of Science (IISc),\\
Bangalore 560012 India\\
Email: \{prashanthis, simmhan\} @iisc.ac.in
}
\date{}

\maketitle

\begin{abstract}
Accelerated edge devices, like Nvidia's Jetson with 1000+ CUDA cores, are increasingly used for DNN training and federated learning, rather than just for inferencing workloads. A unique feature of 
% Nvidia's Jetson GPU-accelerated edges 
these compact devices is their fine-grained control over CPU, GPU, memory frequencies, and active CPU cores, which can limit their power envelope in a constrained setting while throttling the compute performance. Given this vast 10k+ parameter space, selecting a power mode for dynamically arriving training workloads to exploit power--performance trade-offs requires costly profiling for each new workload, or is done \textit{ad hoc}.
% when training on edge devices deployed in the field.
% In this paper, 
We propose \textit{PowerTrain}, a transfer-learning approach to accurately predict the power and time that will be consumed when we train a given DNN workload (model + dataset) using any specified power mode (CPU/GPU/memory frequencies, core-count). It requires a one-time offline profiling of $1000$s of power modes for a reference DNN workload on a single Jetson device (Orin AGX) to build Neural Network (NN) based prediction models for time and power. These NN models are subsequently transferred (retrained) for a new DNN workload, or even a different Jetson device, with minimal additional profiling of just $50$ power modes to make accurate time and power predictions. These are then used to rapidly construct the Pareto front and select the optimal power mode for the new workload, e.g., to minimize training time while meeting a power limit. 
PowerTrain's predictions are robust to new workloads, exhibiting a low MAPE of $<6\%$ for power and $<15\%$ for time on six new training workloads (MobileNet, YOLO, BERT, LSTM, etc.) for up to $4400$ power modes, when transferred from a ResNet reference workload on Orin AGX. It is also resilient when transferred to \repc{an}{two} entirely new Jetson device\addc{s} (Xavier AGX \addc{and Jetson Orin Nano}) with prediction errors of $<14.5\%$ and $<11\%$. These outperform baseline predictions 
by more than $10\%$ and baseline optimizations 
by up to $45\%$ on time and $88\%$ on power. 
\end{abstract}

\section{Introduction}

The growth in domains like autonomous vehicles~\cite{AV_edge} and Internet of Things (IoT) for smart cities~\cite{IoT_edge} are leading to edge computing devices emerging as a first-class computing layer. Edge devices are diverse in their compute capacity and power, ranging from Raspberry Pis to Nvidia Jetsons. Unlike low-end accelerated edges like Google's Coral and Intel's Movidius are designed for Deep Neural Network (DNN) inferencing, Nvidia's Jetson series of GPU accelerated edge devices offer performance that approaches GPU workstations~\cite{matei_europar} with a compact form-factor and low power envelope. E.g., the latest generation Jetson Orin AGX has 2048 Ampere CUDA cores
with performance comparable to an RTX $3080$Ti workstation, but is as small as a paperback novel and has a peak power of under $60$~W. So accelerated edges are competitive candidates for DNN training with power-constraints~\cite{sigmetrics23}.

Edge devices deployed as part of private networks and edge clouds are particularly amenable for use in DNN training.
Training can be done on private edge devices deployed within a single region or over a wide area network, e.g., over data from sensors deployed in factory to detect faults~\cite{CARVALHO2019106024}, from field cameras deployed by scientific collaborations to detect forest fires~\cite{Yang_Lupascu_Meel_2021}, and over video streams from a network of city safety cameras~\cite{Chun_2019_ICCV}. These complement training on pay-as-you-go shared private edge cloud infrastructure that is colocated with content distribution networks (CDNs) and 5G towers~\cite{Silvestre2012CajuAC, Zaw2021EnergyAwareRM}. Besides opportunistically making use of captive compute in private edge cloud systems, training on the edge can avoid the high bandwidth needed to move data to public clouds for training, e.g., from 1000s of traffic cameras
~\cite{kim2022goal}. Lastly, a key motivator for edge training is driven by privacy reasons, e.g., in healthcare and fintech, where data collected on the edge cannot be moved out, even within secure networks due to regulatory requirements. Federated learning~\cite{pmlr_mcmahan_FL,google-sysml} leverages fleets of heterogeneous edge devices in edge cloud systems to train models locally over local data for subsequent aggregation on the cloud across multiple epochs and rounds. 
Such training can happen \textit{recurrently} as part of continuous learning to adapt to data drift~\cite{shmelkov2017incremental,ekya}, and involve \textit{diverse DNN workloads} trained on captive edge devices to meet evolving application needs, e.g., using the same forest cameras to detect forest fires or to classify and count wild animals.

Edge training can also operate under environmental or user constraints. There may be \textit{energy constraints} imposed by power banks on drones or solar-charged batteries in a forest. There can be \textit{power limits} to avoid overheating of poorly ventilated IP-67 enclosures in an industrial site~\cite{sparta-heat-budget-zhang-2022}. There can also be \textit{time constraints}, e.g., to meet a training deadline for an application or limit the billed cost. In federated learning, such time estimates may also guide device selection~\cite{AbelmoniemREFL, ChaiTiFL}.
So, it is important to \textit{configure} the edge accelerator to adapt to these constraints optimally and offer \textit{predictable power, energy and time estimates} for DNN training.

\subsection{Challenges}
A unique feature of Nvidia's Jetson edge devices is their fine-grained control over CPU, GPU and memory frequencies and active CPU cores, enabled by dynamically setting their \textit{power modes}. E.g., the Orin AGX offers 29 CPU frequencies up to 2.2GHz, 13 GPU frequencies up to 1.3GHz, 4 memory frequencies and 12 CPU core-count settings for a total of $\approx 18,000$ possible power modes (Table~\ref{tbl:jetsonspecs}). 
% The previous generation Xavier AGX offer 29k possibilities. 
These can help limit the power envelope in a constrained setting while throttling the performance. However, given this vast parameter space, selecting an \textit{optimal power mode} to make such trade-offs for training workloads that arrive dynamically is non-trivial. Some frequencies span an order of magnitude, and our experiments show that they can have up to $36\times$ impact on DNN training time and $4.3\times$ impact on power usage. For instance, using a low power mode to train the ResNet model on ImageNet data (see \S~\ref{sec:setup} for details) on Orin AGX takes $112$~mins per epoch and consumes about $11.8$~W of power, while increasing it to a higher MAXN power mode reduces the training time to $3.1$~mins but increases the power load to $51.1$~W.

These also vary a lot across \textit{DNN workloads} that may arrive continuously over time, e.g., 
the \bert model on SQuAD dataset using the same MAXN power mode for Orin AGX takes a much longer $68.7$~mins with a power of $57$~W compared to \resnet which only took $3.1$~mins above. 
As expected, these also vary substantially across \textit{Jetson device generations}, which may co-exist in a deployment as new devices are deployed or old ones upgraded in a rolling manner. E.g., the AGX Xavier takes $8.47$~mins per epoch and $36.4$~W to train the \resnet model on ImageNet dataset, as against $3.1$~mins and $51.1$~W for AGX Orin, on the MAXN power mode on both devices.

Hence, a wrong power mode choice can cause high power and performance penalties, violating time and power load constraints or, in the worst case, destroying the device due to overheating. So, accurately estimating the power and time taken for training a given DNN workload on the edge accelerator using a specific power mode is critical~\footnote{Since $energy~(mWh) = power~(mW) \times time~(h)$, we focus on predicting power and time for training, and can derive energy estimates from these.}.

A na\"{i}ve solution of benchmarking all power modes for the device for each new DNN model is intractable. E.g., it takes $16.3~h$ to profile even 25\% of power modes for Orin AGX for the ResNet model on ImageNet data, and this will not scale for every workload, especially if it arrives dynamically and has time constraints. Random or even targeted sampling of some power modes is inaccurate for optimizations, as we show later. Practical deployments may have multiple generations of accelerated devices, amplifying these costs. So, a \textit{principled approach with low overheads and generalizability} is required to build prediction models to estimate training time and power for diverse DNN workloads on edge computing systems.

\subsection{Gaps} 
There is substantial work on optimizing the energy consumption of DNN training on GPU servers~\cite{zeus} and predicting their runtime~\cite{paleo_ICLR}, and power and energy consumption~\cite{NP_ACML}. But these do not translate directly to the edge due to architectural differences, such as the shared memory across CPU and GPU for edge accelerators, and the use of ARM-based rather than x86 CPUs. 
Similarly, several studies have examined energy and performance profiling for edge inferencing rather than training~\cite{edge_config, frqswitching, autoscale-kim-2020}. Some have also explored power modes and frequency scaling for micro-benchmarks on older Jetson architectures~\cite{matei_europar}. However, inferencing workloads have low compute demand compared to training, which can stress all resources of the edge device.

Our previous work focused on characterizing the performance of the previous generation Jetson Xavier AGX and Nano devices for DNN training, but offered only limited preliminary results on actual performance predictions or tuning of power modes~\cite{sigmetrics23}. Nvidia offers a limited set of pre-defined power mode choices ($3$ for Orin AGX apart from the default MAXN) with varying power budgets, but these are too few to allow targeted optimizations. As we show later,

Nvidia's own tool highly overestimates the power load for Orin AGX, causing excess training time when meeting a power budget.
In summary, there is a clear need for modeling, prediction and optimization studies for DNN training workloads on accelerated edges, but none exist. \textit{We address this gap, and offer the first principled modeling and prediction approach of its kind in this work.}

\begin{figure}[t]
\centering
\includegraphics[width=0.6\columnwidth]{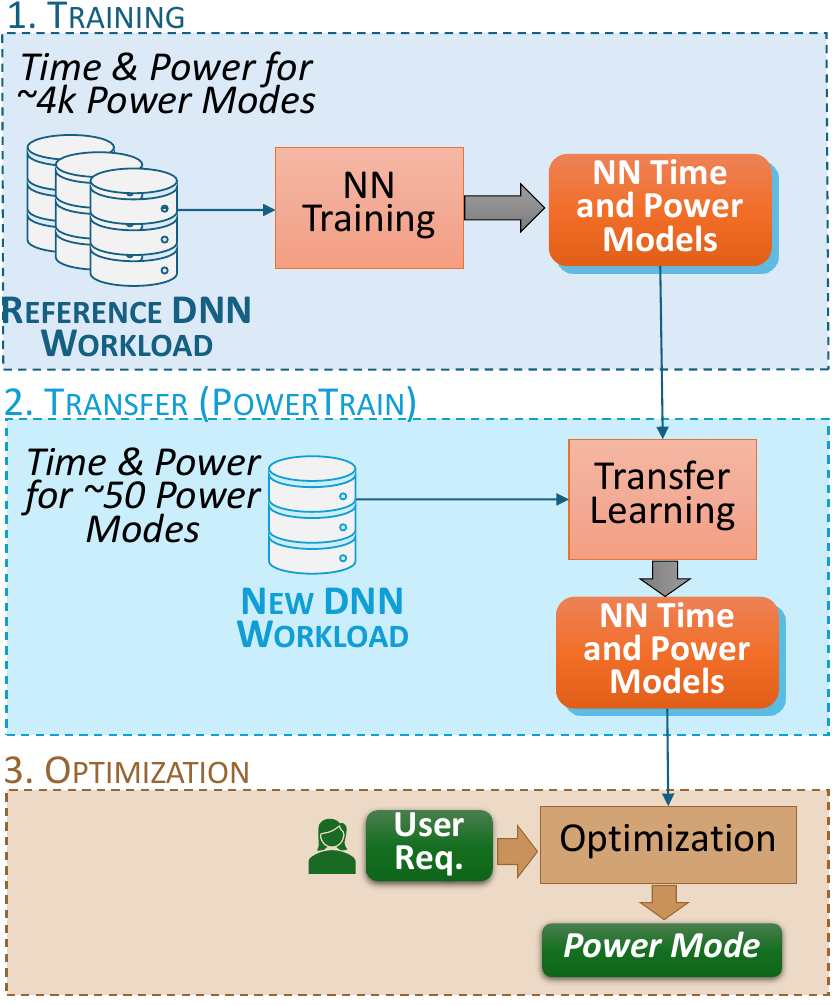}

\caption{Time and Power Prediction Models using PowerTrain: (1) Initial training, (2) Transfer, and (3) Use for Optimization}

\label{fig:workflow:condensed}
\end{figure}

\subsection{Proposed Approach}

We propose two data-driven approaches to address this challenge.
First, as an intelligent baseline, we train two Neural Network (\textbf{NN}) models -- one each for time and power predictions -- over a subset of power modes for the given DNN workload and use this to make predictions for unseen power modes for that same DNN workload. Our second, more flexible approach, \textbf{PowerTrain}, trains such NN models over a large corpus of power modes measured for some reference DNN workload (Figure~\ref{fig:workflow:condensed}). 
When a new DNN workload arrives, it uses transfer learning over the time and power telemetry collected from profiling a small sample of $\approx 50$ power modes for the new workload to retrain updated NN prediction models for it~\footnote{To avoid confusion, we use the term DNN to refer to training workloads, e.g., \mobilenet, \resnet and \yolo and their datasets. We use the term NN to refer to neural network models we train for predicting the time and power for a DNN workload, and PowerTrain (PT) to refer to our specific transfer-learning approach.}.
The PowerTrain prediction models are then used to estimate the expected training time and power for all possible power modes. This is fast and can be used to find the Pareto front~\footnote{The Pareto front is defined as a subset $\widehat{\mathbb{C}}$ from a set of choices $\mathbb{C}=\{c_1, c_2, ...\}$ that affect two optimization (say, minimization) variables $X$ and $Y$ (e.g., power modes that affect time and power). The front contains the set of trade-off choices such that for any given $\widehat{c_i} \rightarrow (x_i,y_i)$ present on the front, there does not exist any other $c_j \rightarrow (x_j,y_j)$ such that $x_j<x_i$ and $y_j<y_i$.} across time and power, and solve a user's optimization goal, e.g. the power mode with the fastest training time for a given power limit, or the lowest power for a given time budget.

As is intuitive, PowerTrain has lower profiling overheads than NN for new workloads and, as we show, offers competitive prediction accuracies for subsequent optimizations
It has a one-time profiling overhead on the reference workload and to train the initial NN prediction models. But the incremental overhead for a new workload is limited to profiling 10s of power modes. Further, PowerTrain generalizes to new DNN workloads, new datasets and, interestingly, to even a new device type, e.g., from a reference model trained on Orin AGX to a new workload running on Xavier AGX.
Using just the NN based prediction models trained from scratch on 10s of power modes for a new workload instead of transfer learning from a pre-trained model is sub-optimal, taking at least twice as much profiling data to achieve similar prediction accuracies as PowerTrain.

\begin{figure*}[t]
  \centering
  \subfloat[Prediction Accuracy, PowerTrain (PT) vs. Nvidia Tool (NV)]{%
    \includegraphics[width=0.33\textwidth]{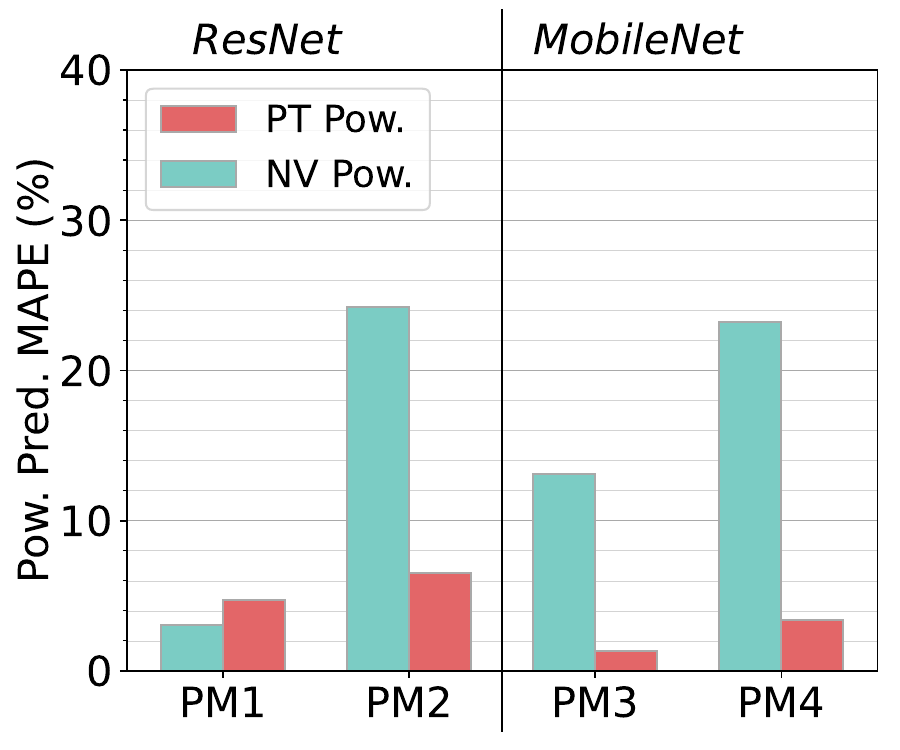}%
    \label{fig:nv:prediction}
  }%
  \hfill
  \subfloat[Optimization Performance, PowerTrain vs. Baseline Models]{%
    \includegraphics[width=0.27\textwidth]{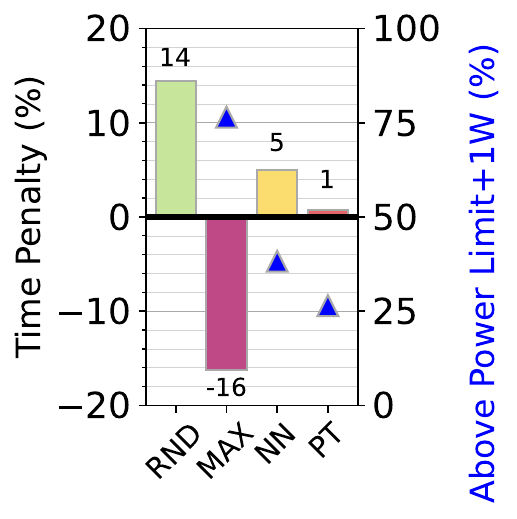}%
    \label{fig:optim:mobnet:time}
  }%
  \hfill
  \subfloat[Optimization Performance, PowerTrain (PT) vs. Nvidia Tool (NV)]{%
    \includegraphics[width=0.36\textwidth]{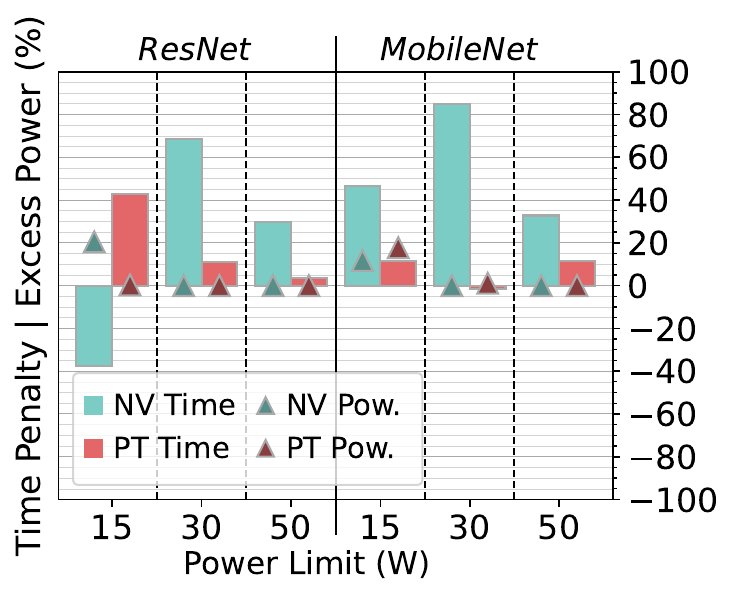}%
    \label{fig:nv:optimization}
  }
  %\vspace{-0.1in}
  \caption{\modc{Representative and comparative results on prediction and optimization. The power modes used in (a) are: \textbf{PM1}:12c/1.65C/0.62G/3.19M, \textbf{PM2}:12c/2.20C/1.23G/3.19M, \textbf{PM3}:10c/1.65C/0.82G/3.19M, \textbf{PM4}:12c/2.20C/1.03G/3.19M}}
 
\end{figure*}

\begin{table*}[t]
\centering
\footnotesize
\caption{Summary of scenarios and solution approaches}
\label{tbl:PracMat}
\begin{tabular}{L{1.5cm}|L{1.5cm}|L{1.4cm}|L{1.3cm}||L{2.5cm}|L{1.7cm}}
\hline

\textbf{Scenario} & \bf {Frequency of training} & \bf {DNN Workload Changes?} & \bf {DNN Training Time} & \bf {Suitable Solution} & \bf {Data Collection Overhead for Solution} \\
\hline\hline
\bf Training once on a large dataset & One time & Never & Few days & Brute force; Profile all power modes and pick the best & $1200$--$1800$~min \\
\hline
\bf Fine-tuning a Model & Occasional & Rare & Few hrs & Neural Network; Profile at least 100 power modes & $20$--$50$~min \\
\hline
\bf Continuous learning & Periodic & Rare & $<1$~hr & PowerTrain; profile 50 power modes and transfer & $10$--$20$~min \\
\hline
\bf Federated learning on edge cloud & Often & Often & Unknown & PowerTrain; profile 50 power modes and transfer & $10$--$20$~min \\
\hline

\end{tabular}
\end{table*}

\subsection{Representative Results against Baselines}
\label{sec:intro:rep}
While a detailed evaluation is presented in \S~\ref{sec:results:predictions} and \S~\ref{sec:results:optimize}, we offer some representative results here to illustrate the comparative benefits of our approach.

\textbf{Predictions.} Nvidia offers a \textit{PowerEstimator} tool (NPE)~\footnote{\url{https://jetson-tools.nvidia.com/powerestimator/}} to estimate the power usage by Orin AGX for a specific power mode. We use this to estimate the power for two diverse power modes (PM$i$) each when training three DNN workloads. 
We also use our PowerTrain (PT) approach, using ResNet as the reference workload on Orin AGX, to predict the power usage for these workloads, and report the Mean Average Percentage Error (MAPE \%) relative to the actual observed power usage in Figure~\ref{fig:nv:prediction}. Other than for PM1 for ResNet, where PT has a slightly larger error (5\%) than NPE (4\%), we give much better predictions in all other cases while NPE consistently overestimates.

\textbf{Optimization.} 
Having predictions for the power and training time for a DNN workload for any given power mode can help users optimize the system to meet different goals. E.g., in \S~\ref{sec:results:optimize}, we find the power mode for a DNN training workload such that the power remains within a budget while the training time is minimized.
Nvidia recommends $3$ default power modes~\footnote{\url{https://docs.nvidia.com/jetson/archives/r35.1/DeveloperGuide/text/SD/PlatformPowerAndPerformance/JetsonOrinNxSeriesAndJetsonAgxOrinSeries.html}} along with their power budgets for users to choose from: $15$~W, $30$~W and $50$~W. We use the best of these three to meet an optimization goal, defined as falling within a power limit while minimizing training time.
We also solve this using our PowerTrain prediction models to discover a Pareto front (described in \S~\ref{sec:results:optimize}) to select a custom power mode. We compare the observed training time for these two solutions against the ground-truth optimal found through brute force for these 3 power limits. Figure~\ref{fig:nv:optimization} shows that PowerTrain (PT) has the fewest \% of solutions that exceed the optimal for 5 out of 6 cases (except \resnet 15W) compared to Nvidia's suggestions (NV), while falling within the power limits in most cases.

Lastly, in Figure~\ref{fig:optim:mobnet:time}, we see that the PT-based optimization also outperforms simpler baselines like choosing the default MAXN power mode and doing random (RND) profiling of 50 power modes to build a partial Pareto and selecting the best, and our better baseline using a Neural Network (NN) trained on 50 power profile samples.
For a set of optimization problems with power limits varying from $17$--$50W$, PT has the lowest time penalty of $1\%$ in excess of the optimal while also having the lowest percentage ($26.5\%)$ that exceeds the power limit by over $1W$ above the threshold.  \addc{This low time penalty is especially beneficial when doing long training runs over several epochs. For instance, we see that a typical training of YOLO takes $200$ epochs to converge, lasting almost $49$ hours on the Orin. A $12\%$ time benefit achieved by PT accrues over all epochs and the training can complete $5.88$~hours faster. Similarly, \mobilenet takes $148$ epochs ($50$ hours) to converge, with time reduction of $6.5$~hours when optimized using PT.}

These exemplars motivate the need for and illustrate the success of our robust PowerTrain prediction model to support power mode tuning for training DNN workloads on Jetson edge devices.

\subsection{Discussion}
It is useful to understand the scenarios under which these different approaches, including baselines and our proposed ones, are best suited for.
In Table~\ref{tbl:PracMat}, we outline four potential application and deployment scenarios, their characteristics and the solution approach is most applicable. These estimates are based on our detailed experiments later reported in Section~\ref{sec:results:predictions}.

For instance, there may be a one-time training workload submitted to an edge service that runs new and large DNN model on a large corpus of user data. Since the training is time-consuming, running into days, having the perfect power mode is helpful even if the profiling time is costly. Here, a brute force approach may be recommended since the data collection time is approximately a day, which is well within the runtime of the workload. This kind of training workload is uncommon on the edge, and this approach is therefore impractical. 

Fine-tuning a pre-trained DNN on a smaller dataset is a more common edge usecase (e.g., model personalization on a private edge device) where the DNN training time is typically a few hours. In this scenario, the DNN workload seldom changes and our NN approach using profiling over 100s of power models is feasible because of the longer training time. 

Continuous learning is another common scenario on edge devices where the same DNN is regularly retrained with small batches of new incoming data to address data drift. Since the training workload runs for a shorter time here, PowerTrain is the best approach since it has a smaller data collection time. Further, when the edge device is part of a federated learning setup or a private edge-cloud supporting multiple applications, the device could be assigned any DNN training workload, submitted often and with an unknown training duration. In such a case, it is best to use PT to minimize data collection time for the new workload and ensure that the optimization criteria are met before the workload changes.

\subsection{Contributions} 
We make the following specific contributions in this article:

\begin{enumerate}[leftmargin=0.5cm,itemindent=0cm,labelwidth=0.5cm,labelsep=0cm,align=left]
 \item We develop a Neural Network (NN) based prediction approach to \textit{estimate the time and the power} for DNN training workloads on the Nvidia Jetson Orin AGX for any given power mode (\S~\ref{sec:emp_ml}) using a large profiling corpus. This is extended into PowerTrain (PT), 
 a \textit{Transfer Learning} based approach that greatly reduces the online profiling overheads for any new DNN workload on Orin or another Jetson device. 
 
 \item We comprehensively validate NN and PT for accuracy and generalizability across 7 different DNN workloads and\repc{a}{three} different Jetson edge device\addc{s (Orin AGX, Xavier AGX and Orin Nano)}. \addc{We also validate PT predictions on 3 different training minibatch sizes across 2 workloads.} (\S~\ref{sec:results:predictions}).
    \item Finally, we apply PT predictions to optimize the training time and power limit trade-offs for several DNN workloads (\S~\ref{sec:results:optimize}), and demonstrate that it out-performs other baselines, including NN.
\end{enumerate}

This modeling work leverages a prior in-depth study that we conducted on the effect of various factors, including power modes, on the Jetson devices for DNN training~\cite{sigmetrics23}. There, the focus was on the characterization and as a potential use-case of that study, we had proposed a simple linear regression technique (Section 5.7 in ~\cite{sigmetrics23}) for predicting the DNN training time and energy on the Jetson devices, validated on 10 power models for one Jetson device, with mixed results. 
The present article focuses entirely on predicting the time and power for DNN workloads on Jetsons, proposing a wholly different, novel NN-based prediction technique along with the PowerTrain transfer learning approach for fast retraining and robust estimates. We further present optimization results as a case study of using our prediction models. All these contributions are novel, with no more than 20\% conceptual overlap with the prior work.

This article is a pre-print of the article that has appeared in Elsevier's FGCS~\cite{fgcs}.

\begin{comment}
\end{comment}

\section{Experiment Setup}%\Note{2.5 pg}
\label{sec:setup}

We first offer a detailed description of the DNN workload and training configurations to highlight the rigor of our profiling setup, the diversity of our workloads, and to allow reproducibility.

\subsection{Choice of hardware platform, applications, etc.}
Nvidia Jetsons are arguably the most popular high-end accelerated edge devices~\cite{sigmetrics23} and are leading the MLPerf benchmarks~\cite{mlperf_inf}. Hence, we develop our models and validate them on Jetson devices. \addc{As a point of reference to compare the compute power of these Nvidia Jetson devices, we run our $5$ training workloads on $3$ additional device types (server GPU, workstation GPU, Raspberry Pi 5) and report their training times against the Nvidia Jetson Orin used in our experiments. These are reported in the Appendix in Figure A.14. The Raspberry Pi 5 is two orders of magnitude slower than the Orin, has only CPU cores for compute and is unlikely to be used for training heavy models. It also has a much smaller power footprint (1-10W), making power optimizations less relevant. On the other hand, server GPUs come with pre-defined power limit knobs that can internally trigger frequency scaling automatically to stay within a power limit~\cite{zeus}, and do not need methods proposed by us. The Orin offers an interesting combination of near sever-grade compute power with a lot of power modes to tradeoff power and performance.}
That said, these techniques can later be extended to other hardware platforms which offer a choice of power mode trade-offs.

Since \textit{energy} can be derived as a function of \textit{power} and \textit{time}, we focus on these two component metrics. Further, we focus on time taken for a minibatch of training as this is the practical unit of training, and the time taken for an epoch can be derived by multiplying the minibatch time with the number of minibatches in one epoch of the training data. We focus on DNN training workloads given their need for high compute and the growing prevalence of DNN training on the edge, e.g., federated learning. Other non-DNN workloads can be considered as future work.

\begin{table}[t]
\centering
\footnotesize
\caption{Specifications of NVIDIA Jetson Orin AGX, Xavier AGX, and Orin Nano devkits used in evaluations}
\label{tbl:jetsonspecs}

\begin{tabular}{l|r|r|r}
\hline
\textbf{Feature} & \textbf{Orin AGX} & \textbf{Xavier AGX} & \textbf{Orin Nano} \\
\hline\hline
CPU Architecture & ARM A78AE & ARM Carmel & ARM A78AE \\ 
\noalign{\global\arrayrulewidth=0.1pt}\arrayrulecolor{lightgray}\hline
\noalign{\global\arrayrulewidth=0.4pt}\arrayrulecolor{black}
\# CPU Cores & 12 & 8 & 6 \\
\noalign{\global\arrayrulewidth=0.1pt}\arrayrulecolor{lightgray}\hline
\noalign{\global\arrayrulewidth=0.4pt}\arrayrulecolor{black}
GPU Architecture & Ampere & Volta & Ampere \\
\noalign{\global\arrayrulewidth=0.1pt}\arrayrulecolor{lightgray}\hline
\noalign{\global\arrayrulewidth=0.4pt}\arrayrulecolor{black}
\# CUDA/Tensor Cores & 2048/64 & 512/64 & 1024/32 \\
\noalign{\global\arrayrulewidth=0.1pt}\arrayrulecolor{lightgray}\hline
\noalign{\global\arrayrulewidth=0.4pt}\arrayrulecolor{black}
RAM (GB)/Type & 32/LPDDR5 & 32/LPDDR4 & 8/LPDDR5 \\ 
\noalign{\global\arrayrulewidth=0.1pt}\arrayrulecolor{lightgray}\hline
\noalign{\global\arrayrulewidth=0.4pt}\arrayrulecolor{black}
Peak Power (W) & 60 & 65 & 15 \\ 
\noalign{\global\arrayrulewidth=0.1pt}\arrayrulecolor{lightgray}\hline
\noalign{\global\arrayrulewidth=0.4pt}\arrayrulecolor{black}
Form factor (mm) & 110 x 110 x 72 & 105 x 105 x 65 & 100 x 79 x 21 \\ 
\noalign{\global\arrayrulewidth=0.1pt}\arrayrulecolor{lightgray}\hline
\noalign{\global\arrayrulewidth=0.4pt}\arrayrulecolor{black}
Price (USD) & \$1999 & \$999 & \$499 \\ 
\hline
\end{tabular}

\vspace{0.5cm}

\begin{tabular}{L{4cm}|r|r|r}
\hline
\textbf{Features that vary across Power Modes} & \textbf{Orin AGX} & \textbf{Xavier AGX} & \textbf{Orin Nano} \\
\hline\hline
CPU core counts & 1..12 & 1..8 & 1..6 \\ 
\noalign{\global\arrayrulewidth=0.1pt}\arrayrulecolor{lightgray}\hline
\noalign{\global\arrayrulewidth=0.4pt}\arrayrulecolor{black}
\# CPU freqs. & 29 & 29 & 20 \\ 
\noalign{\global\arrayrulewidth=0.1pt}\arrayrulecolor{lightgray}\hline
\noalign{\global\arrayrulewidth=0.4pt}\arrayrulecolor{black}
Max. CPU Freq. (MHz) & 2200 & 2265 & 1500 \\ 
\noalign{\global\arrayrulewidth=0.1pt}\arrayrulecolor{lightgray}\hline
\noalign{\global\arrayrulewidth=0.4pt}\arrayrulecolor{black}
\# GPU freqs. & 13 & 14 & 5 \\ 
\noalign{\global\arrayrulewidth=0.1pt}\arrayrulecolor{lightgray}\hline
\noalign{\global\arrayrulewidth=0.4pt}\arrayrulecolor{black}
Max. GPU Freq. (MHz) & 1300 & 1377 & 625 \\ 
\noalign{\global\arrayrulewidth=0.1pt}\arrayrulecolor{lightgray}\hline
\noalign{\global\arrayrulewidth=0.4pt}\arrayrulecolor{black}
\# Mem freqs & 4 & 9 & 3 \\ 
\noalign{\global\arrayrulewidth=0.1pt}\arrayrulecolor{lightgray}\hline
\noalign{\global\arrayrulewidth=0.4pt}\arrayrulecolor{black}
Max. Mem. Freq. (MHz) & 3200 & 2133 & 2133 \\ 
\noalign{\global\arrayrulewidth=0.1pt}\arrayrulecolor{lightgray}\hline
\noalign{\global\arrayrulewidth=0.4pt}\arrayrulecolor{black}
\# Power modes & 18,096 & 29,232 & 1,800 \\ 
\hline
\end{tabular}
% \vspace{-0.15in}
\end{table}

\begin{table*}[t]
\centering
\setlength{\tabcolsep}{1.5pt}
\footnotesize
% \vspace{-0.1in}
\caption{DNN workloads and Datasets used in Experiments. All models are trained with minibatch size of 16. Estimated epoch training time (mins) for MAXN fastest power mode on Orin AGX is given as reference.}
\label{tbl:modeldataset}

\begin{tabular}
{L{1.8cm}|L{2cm}|R{1cm}R{1.2cm}R{1.2cm}||R{2cm}|R{0.5cm}R{1.1cm}|R{0.6cm}}
\hline
 \bf{Task} & \bf{Model} & \bf{\# Layers} & \bf{\# Params} & \bf{FLOPs}$^\dagger$ & \bf{Dataset} & \bf{\# Samples} & \bf{Size} & \bf Epoch Time (min) \\

  \hline\hline

 Image classification & \textbf{{\mobilenet}v3~\cite{mobilenet}}& 20  & $5.5M$ & $225.4M$ & \textbf{GLD23k~\cite{tensorflow_gld23k}} & 
 % $23,080$  
 $23k$  
 & $2.8GB$ & $2.3$\\
  \hline
 Image classification & \textbf{\resnet-18~\cite{resnet}}& 18  & $11.7M$  & $1.8G$   
 & \textbf{ImageNet Val.~\cite{imagenet}} & 
 % $50,000$ 
 $50k$ 
 & $6.7GB$ & $3$\\
  \hline
 Object detection &\textbf{\yolo-v8n~\cite{yolo}} & 53 &  $3.2M$ &  $8.7G$   &  \textbf{COCO minitrain~\cite{coco_minitrain}} &  
 
 $25k$
 & $3.9GB$ & $4.9$\\

  \midrule
  \midrule
 Question answering & \textbf{BERT base~\cite{bert}}& 12 & $110M$  & $11.5T$   
 & \textbf{SQuAD~\cite{squad}} & 
 
 $70k$ 
 & $40MB$ & $68.6$\\
    \hline
 Next word prediction & \textbf{LSTM~\cite{LSTM}}& 2  & $8.6M$  & $3.9G$   
 & \textbf{Wikitext~\cite{wikitext}} & 
 
 $36k$ 
 & $17.8MB$ & $0.4$\\

\hline
\multicolumn{8}{L{12cm}}{$^\dagger$~ As per the typical practice, FLOPs reported correspond to a forward pass with minibatch size 1. }

\end{tabular}

\end{table*}

\subsection{Hardware and Settings}

Nvidia Jetson Orin AGX~\cite{Orin} is the latest generation of the Jetson edge accelerators, released in March 2023.
The specifications of the development kit used in this study is given in Table~\ref{tbl:jetsonspecs}. The Orin supports several custom power modes, as described earlier and also shown in the table. The Orin devkit features INA3221 power sensors, from which we read the present power consumption of the device during our experiments. 
We set the fan to maximum speed to avoid any thermal throttling effects during our experiments.
We disable Dynamic Voltage and Frequency Scaling (DVFS) so that the frequencies remain static at the value we have configured them to. Beside the GPU CUDA cores, the Orin also has two special purpose accelerators, DLA and PVA, which we do not use and leave in their default states and turned off. We also use the Nvidia Jetson Xavier AGX~\cite{Xavier} developer kit, the predecessor of Orin AGX, \addc{and the Jetson Orin Nano~\cite{Orin-nano}, a less powerful device in the same generation of Orin AGX} for generalizability experiments in a later section.\repc{We refer to the two devices as Orin and Xavier respectively.}{We refer to the three devices as Orin, Xavier and Nano, respectively.}

\subsection{Training Framework and DNN Workloads}
Orin runs Ubuntu 20.04 LTS and L4T kernel version $5.10.65$. We configure it with Nvidia JetPack version $5.0.1$, CUDA v$11.4$, PyTorch $1.12$ and torchvision v$0.13$.

We select as our default workloads three popular computer vision DNN architectures and datasets that can train within the available resources of the 
Orin (Table~\ref{tbl:modeldataset}): \textit{\resnet-18} and \textit{\mobilenet v3} that perform image classification, and \textit{\yolo v8n} that does object detection task, coupled with training datasets to form the workloads. \mobilenet is a lightweight vision model optimized for smartphone applications. Our workload trains \mobilenet using images from the \textit{Google Landmarks Dataset v2 (GLD-23k)}, which consists of 23,080 photos of both human-made and natural landmarks, categorized into 203 different classes. \resnet is a family of Convolutional Neural Networks (CNNs) used for image classification, featuring residual blocks and skip connections. We train \resnet-18 on the validation subset of \textit{ImageNet}, which consists of 50,000 images and occupies 6.7GB on disk. 

\textit{\yolo} v8 is the latest iteration in the popular ``You Only Look Once'' series of real-time object detectors. We use \yolo v8n, the smallest of its family, to train on a subset of the MS COCO dataset, \textit{COCO minitrain}, which has 25,000 images which take up 3.9GB on disk. 
In addition, we also use \textit{\bert} and \textit{\lstm} DNN architectures with the SQuAD and Wikitext training datasets, used for query-response and next-word prediction, respectively, to explore the generalizability of our methods to other DNNs. All of these are representative of typical edge and federated learning workloads~\cite{AbelmoniemREFL, jallepalli_federated_yolo, tian_fedbert}
, and provide a wide diversity in DNN architectures (CNN, LSTM, Transformer), dataset sizes (17.8MB--6.7GB) and computational requirements (3.2M--110M parameters, 225M--11.5T FLOPS). 

\textit{PyTorch} is used for training the DNNs and its Dataloader module is used to fetch and pre-process the data samples. A minibatch of the training data forms the smallest unit of data fetch and training. The $num\_workers$ flag is used to set the number of fetch and pre-process workers. When $num\_workers=0$, a single process handles all data transfer and GPU compute operations, with no pipelining. When the $num\_workers$ flag is set to $n \geq 1$, $n$  processes are initiated for fetching and pre-processing, with
a separate process carrying out GPU computation on each pre-processed minibatch sequentially. By default, we use a minibatch size of $16$ samples during training and $num\_workers=4$ ~\footnote{The \yolo model version has a bug for $num\_workers=4$ setting, which is fixed in a later version of PyTorch but not available for our Jetson JetPack version (\url{https://github.com/pytorch/pytorch/issues/48709}).
Hence, we use $num\_workers=0$. This is valid, but causes \yolo training to have GPU stalls as the main process is responsible both for the data loading and the compute.}.
We run an iteration of the DataLoader to pre-fetch all data to memory before starting the workload to avoid any disk fetch overheads during our data collection.

\subsection{Profiling Setup and Metrics}
During profiling of the DNN workload training, we sample the current power of the board in milliWatt (mW) every 1s using the \textit{jtop} library, which is a wrapper around \textit{tegrastats}. This gives the Jetson module's power as reported by the power sensors and not the overall power drawn by the carrier board. This scopes the power to the key resource components used in the DNN training, and power drawn by other peripherals like USB ports, etc. are avoided, which in any case have been reported to be negligible~\cite{sigmetrics23}.

We add instrumentation to the PyTorch code to measure the execution time for every minibatch in milliseconds (ms). This is done using \texttt{torch.cuda.event} with the \texttt{synchronize} option to accurately capture time spent on the GPU. We experimentally verify that our profiling overhead is minimal and does not affect the runtime of the workload. Before we start the workload, we pre-cache the dataset into the page cache to avoid part of the data being fetched from memory and the rest from disk. This gives consistent minibatch training performance.

%% ===================================================================================================

\subsection{Data Collection from Profiling Power Modes}
\label{sec:exp:data}

Our prediction models use as input the profiling information collected during DNN workload training for a specific power mode of the device.
When profiling each power mode, we train the candidate DNN workload for $\approx 40$ minibatches, and record the training time per minibatch and the sampled power usage during the training period. \addc{We conducted a sensitivity study by varying the number of minibatches used when profiling each power mode and evaluating its impact on the accuracy of the trained time and power prediction models. There is minimal impact on the pre-training and fine-tuning accuracies when varying from $10$--$40$ minibatches (e.g., MAPE range of $10.0$--$14.9\%$ for time prediction, $3.2$--$4.4\%$ for power prediction during pre-training) since the training time and power are remain stable across minibatch samples. However, we notice that with fewer minibatches and at faster power modes, the power profiling is unable to collect any telemetry since the sampling interval is $1s$ and the training of all the minibatches completes within this interval. As a result, we conservatively pick $40$ minibatches to train over during profiling. }

There is negligible variation in the time or power measured across the minibatches during DNN training. However, we note that the very first minibatch's training takes a long time to train, possibly due to PyTorch performing an internal profiling to select the best possible kernel implementation in this phase. Therefore, we discard the first minibatch profiling entry. Additionally, we notice that the power measurements when training using a new power mode take $2$--$3$s to stabilize. So, we use a sliding window logic to detect when the power stabilizes, and use the profiling entries subsequent to that while ensuring 40 ``clean'' minibatch training samples are collected per power mode.
 
For the three default workloads, \resnet, \mobilenet and \yolo, we collect and assemble a large corpus of ground truth power mode profiling data to help us with prediction model training and validation of their performance. The Orin power modes span $29$ CPU frequencies, $13$ GPU frequencies, $4$ memory frequencies and $12$ CPU cores for a total of $18,096$ possible power modes~\footnote{Interestingly, the number and actual frequencies available change with the Board Support Package (BSP) version. We also noticed that the same Orin device, but with a different JetPack version, had a different set of frequencies available. Nvidia's usually active forums do not have any comments related to this.}. From these, we choose $4,368$ power modes for profiling that are uniformly distributed through the solution space; these include all combinations of GPU ($13$) and memory ($4$) frequencies, even number of CPU cores ($6$), and every alternate CPU frequency that is available, excluding the two slowest ones ($14$)~\footnote{During our experiments, we found that the Jetson device only supports changing from higher to lower CPU and GPU frequencies. Other changes require reboots, and cause an error otherwise. We came up with an ordering of the power modes that satisfies this requirement.}. The data collection time depends on the power mode chosen, e.g., a lower configuration will take longer to profile $\approx 40$ minibatches of training, and we report these times as part of our results.

This large corpus of profiling data for different power modes for diverse DNN training workloads helps us rigorously evaluate our prediction models. This dataset collected is available at \texttt{Anonymized} and in itself forms a rich community asset.

\section{ML-driven Modeling and Prediction} %\Note{2.5pg}
\label{sec:emp_ml}

\begin{figure*}[t]
\centering
\includegraphics[width=1\textwidth]{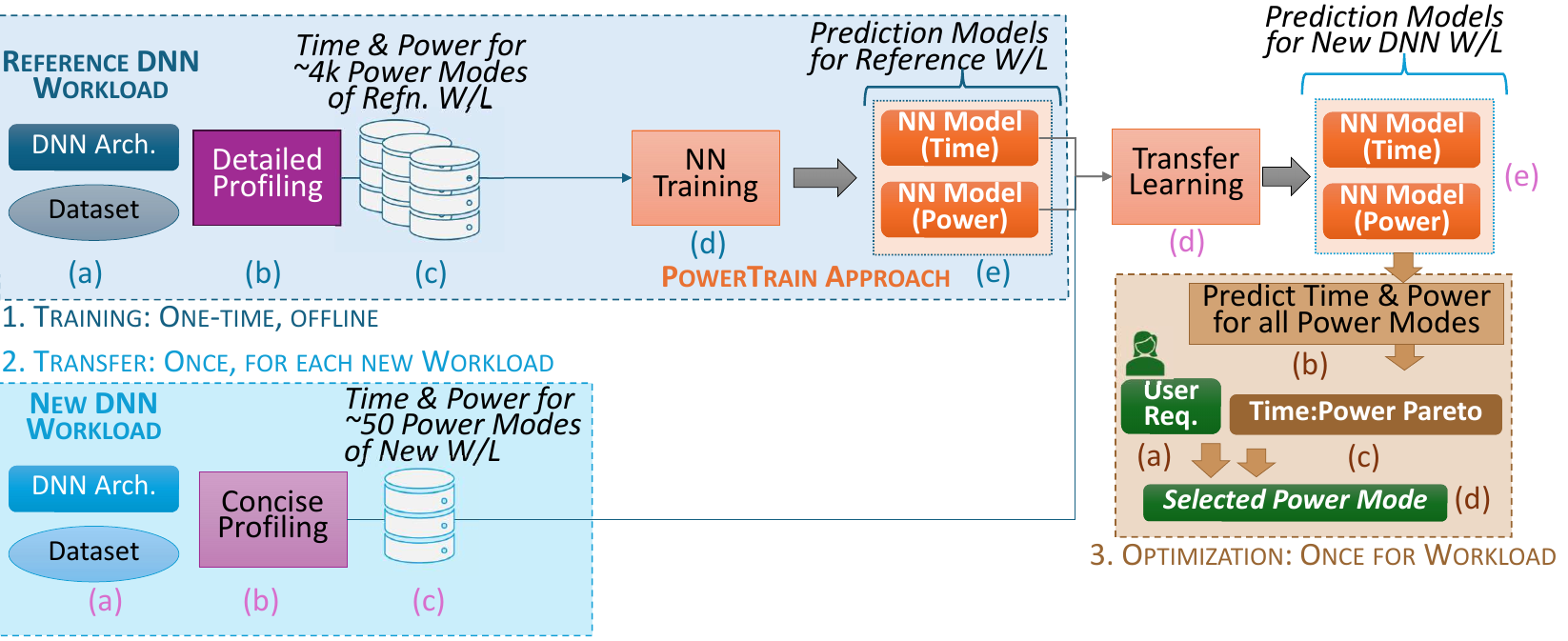}
\caption{\modc{Time and Power Prediction Models: (1) Initial training, (2) Transfer, and (3) Use for Optimization}}

\label{fig:workflow}

\end{figure*}

Here, we propose two data-driven approaches based on ML-based modeling to predict the power and training time per minibatch for a DNN training workload. 

Our proposed methods were arrived at after exploring and eliminating other simpler prediction methods such as linear regression, random forest and decision trees, which did not perform well: DNN workload performance appeared inherently non-linear and linear regression was inaccurate; and random forest and decision trees suffered from overfitting. We also explored using a Multi-Layer Perceptron (MLP) to predict time and power. While this performed well with very few samples, our eventual solution based on Neural Networks (PT) outperforms it with a slightly larger number of samples but with the key added benefit of being able to transfer the learning to a new workload. 

A key aspect that we consider here is the \textit{overhead} for training the prediction models for a new DNN workload and, potentially, new hardware as well. This overhead is manifest primarily in the form of \textit{profiling time} since the actual time to perform prediction model training itself is negligible, taking $15$~minutes at most. NN models tend to be more accurate with more training samples, yet collecting profiling data for a growing number of power modes can take a linearly longer time. While the effort spent profiling can be reused as part of the actual training activity of the DNN workload (i.e., it is not wasted even from the user's perspective), the benefits of constructing the prediction model and performing optimizations on it are subdued if this data collection time starts being a large part of the multi-epoch DNN workload training time.

\subsection{Neural Network~(NN) Prediction Model per Workload}

Neural Networks (NN)~\cite{mcculloch1943logical} consist of 
multiple layers of interconnected neurons, which allow it to learn complex, non-linear relationships in the data. Neural networks usually contain one input layer, one or more hidden layers, and an output layer.
We first develop simple NN architectures to predict the training time and the power for a given power mode on an edge accelerator, for a specific DNN workload.

\begin{table}[t]
\centering
\begin{minipage}[t]{.5\textwidth}
\centering
\begin{tabular}{L{3.3cm}|R{3.2cm}}
\hline
\textbf{Feature} & \textbf{Value} \\
\hline\hline
Layers & 4 (Dense)\\ \hline
Dropouts & After layers 1 and 2 \\ \hline
No of neurons & 256, 128, 64, 1 \\ \hline
Activation & ReLu x 3, Linear \\ \hline
Optimizer & Adam \\ \hline
Loss function & MSE \\ \hline
Learning rate & 0.001 \\ \hline
Training epochs & 100 \\ 
\hline \hline
Profiling minibatches & 40 \\ \hline
Power modes (Ref) & 4,368 \\ \hline
Power modes (TL) & 50 \\ \hline
\end{tabular}
\captionof{table}{\addc{NN hyperparameters}}
\label{tbl:hyper}
\end{minipage}
\end{table}

\begin{figure*}[t]
 \centering
\includegraphics[width=1\textwidth]{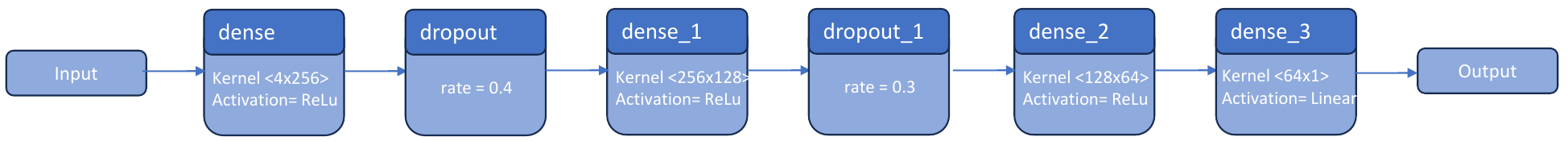}

\caption{\addc{NN Architecture}}

\label{fig:nn_arch}

\end{figure*}

Our \textit{NN architecture} \addc{(Figure~\ref{fig:nn_arch})} has $4$ dense layers with $256$, $128$, $64$ and $1$ neurons, respectively. We use the ReLu activation function for the first $3$ dense layers, and a linear activation function for the final layer. This was arrived at based on a grid search for hyper-parameters using \textit{sklearn} library's GridSearchCV. 
The input feature vector consists of the configuration for a power mode: CPU cores, CPU frequency, GPU frequency and memory frequency, while the output layer returns the predicted training time per minibatch or power. Each input feature is normalized to a value between 0.0 to 1.0 using the \textit{sklearn} library's StandardScaler to ensure better model generalization and faster convergence by providing consistent scales for all inputs.
Adam is used as the optimizer, with a learning rate of $0.001$ and the Mean Squared Error (MSE) serves as the loss function. We also introduce two dropout layers after the first and second dense layers to avoid overfitting. By default, we train the NN for 100 epochs and verify convergence. We use model checkpointing to save the best weights (i.e., model with the least validation loss) seen during training and use it as the final model. \addc{The hyperparameters are listed in Table~\ref{tbl:hyper}.}

We take two alternatives for providing the training data to the NN: (1) \textit{All} gives it the full profiling data for a DNN workload using a 90:10 split of training and test, which is about $4.4$k samples for our Orin AGX for the default DNN workloads. (2) The second provides a much smaller number of training data, from 10 to 100 uniformly sampled from the $4.4k$, again using a 90:10 split.
Clearly, the first approach is time-consuming to collect the profiling data, e.g., taking over $14h$ for ResNet, but can give better prediction accuracy. In contrast, a single epoch of training for ResNet using the MAXN faster power mode takes $3$~mins to train, with a typical training lasting for $120$ epochs, or $6$~h. The latter sampling approach takes a much smaller time, running into $10$s of minutes, but can suffer from lower accuracy. We explore these trade-offs in the evaluation.

\subsection{PowerTrain: Generalizable Transfer Learning Model}

\begin{figure}[t]
\centering
\includegraphics[width=0.75\columnwidth]{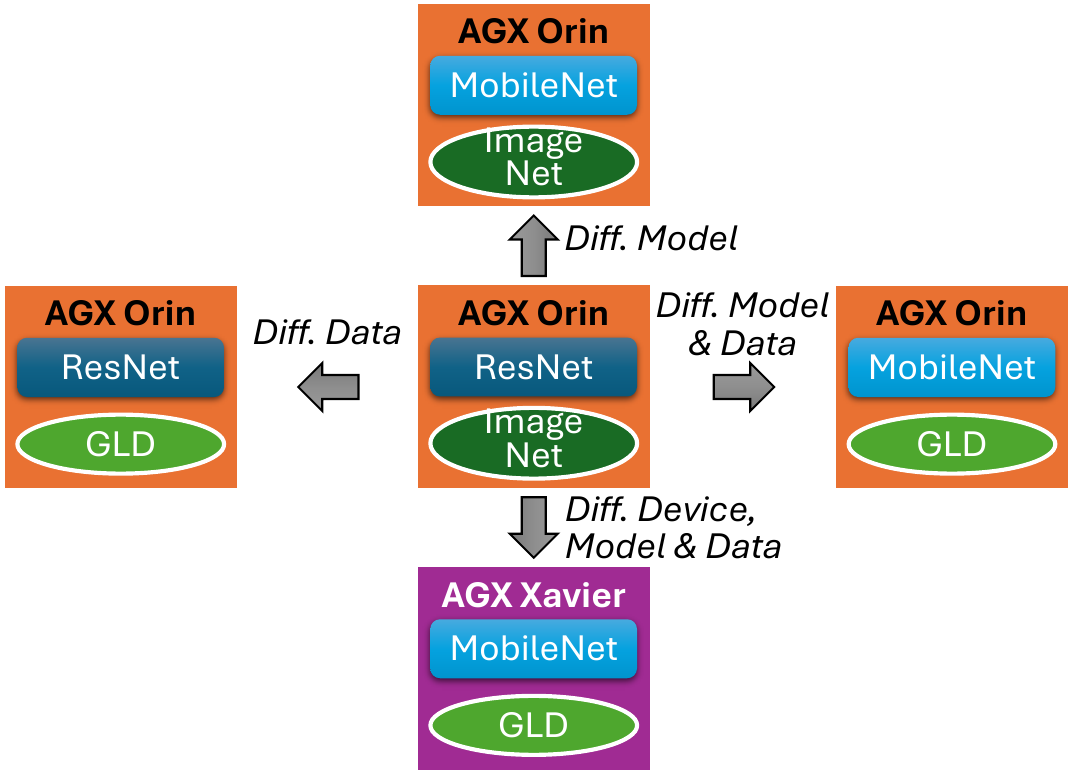}
\caption{Generalizability of PowerTrain.}

\label{fig:workflow:TLmnemonic}

\end{figure}

While the NN approach is simple, it needs to be retrained for every new DNN workload we wish to run on the edge device. This requires fresh data collection overheads for different power modes of the new DNN workload.
To overcome this limitation, we propose \textit{PowerTrain}, which leverages Transfer Learning (TL) from one NN to another to offer a lower overhead generalizable approach.

Transfer learning~\cite{bozinovski1976-original-transfer-learning, bozinovski2020reminder} is a popular technique in deep learning where a model developed for one task is applied to a different but related task. The core idea behind this is that a model trained on a large and generalized dataset can perform well on a related problem, but with significantly less specialized data and training time.

Adopting this,\repc{PowerTrain first bootstraps a reference NN model each for time and power using the architecture described above for any one DNN workload on a large set of profiling data (see Figure 3)}{PowerTrain uses a reference DNN workload (1(a) in Figure~\ref{fig:workflow}), performs detailed profiling (1(b)) to collect a large set of profiling data (1(c)) which is then used to bootstrap a reference NN model each for time and power (1(e)) using the architecture described above.} This is done once, offline, and serves as the reference NN to transfer from. 
Subsequently, when a new DNN workload arrives, we modify this reference NN slightly by removing the last dense layer and adding a fresh layer. 
\repc{We then fine-tune the reference NN model by training it on a limited set of profiling data for the new DNN workload.}{We then take the new DNN workload (2(a) in Figure~\ref{fig:workflow}), perform limited profiling (2(b)), and use this time and power data (2(c)) to fine-tune the reference NN model by performing Transfer Learning (2(d)). This again generates NN models, one each for time and power (2(e)).} The intuition is to retain and utilize the representations learned in the internal layers of the neural network for the reference DNN workload and only change the final output layer in accordance with the new DNN workload's profiling behavior. We do this for both time and power prediction models.

As discussed later (\S~\ref{sec:results:predictions}), we train the reference NN for one of the three default DNN workloads using the full corpus of $4.4k$ power modes as training data. 
We then transfer this to other DNN workloads, but using only a small number power modes for profiling, which serve as training data for transfer learning. We later discuss the choice of this reference DNN workload. 

\addc{We test the impact of the number of power modes used in training the reference NN model, increasing it from $500$ to $4368$. We do not observe any significant difference in MAPEs when predicting time or power for the other power modes of the reference model, or for the predictions returned by the transfer-learned models. Since training the reference model is a one-time profiling activity, we use the full set of $4.4k$ power modes.} We also use different counts of randomly sampled power modes, from 10 till 100, for transfer learning and evaluate the trade-off between data collection overheads and the accuracy of the retrained models for time and power prediction. The actual time for retraining itself remains in the order of $<10$~mins.

PowerTrain can be generalized across DNN architectures and datasets (which together form a new workload), on the same device the reference model was trained on. We also attempt to transfer it to a DNN workload running on a different Jetson device, Xavier AGX.
Specifically, we evaluate PowerTrain for three generalization scenarios (Figure~\ref{fig:workflow:TLmnemonic}): (1) Same DNN architecture or dataset as the reference DNN workload, (2) Unseen DNN architecture and dataset, \delc{and}(3) Unseen device from a different generation\addc{, (4) Unseen device from the same generation, and (5) Unseen training minibatch sizes}. Results from these are presented next.

\section{Prediction Results}%\Note{4 pgs}
\label{sec:results:predictions}

In this section, we present a detailed evaluation of PowerTrain (PT) for time and power predictions, and compare it with the Neural Network (NN) baseline using large and small sampling. These are evaluated on their Mean Absolute Percentage Error (MAPE\%) relative to the ground truth time taken and power for a power mode based on actual measurement, and also on the overheads for data collection. We also evaluate the generalizability of PowerTrain to new DNN workloads and devices.

The PowerTrain reference model and NN large sampling (\textit{All}) are both trained on 90\% of the profiling data from all $4386$ power modes, with $40$ minibatches of telemetry entries available for each power mode. We also use \textit{smaller samples} for NN training and for transfer learning in PowerTrain. Here, we randomly select between $10$--$100$ power modes from $4386$, with $40$ minibatch entries each. Since power data is recorded every $1s$, each power mode has a different number of power samples depending on the duration for profiling $40$ minibatches. 
We use the maximum length of power samples as our entry count, and replicate power mode minibatch entries in case fewer are available to ensure the same number of training entries. 
By default, the \textit{validation} of these models is done on all $4386$ power modes.
We perform $10$ training and validation runs for every configuration and report the median results for these along with $Q1$--$Q3$ quartile range values as whiskers.

Training the full NN for the reference DNN takes around $15$mins on a RTX 3090 GPU, but this can be done offline and the model weights saved. 
The training time for the NN model on $10$--$100$ of samples takes a few seconds. Similarly, PowerTrain takes under $30s$ for transfer learning.

\begin{figure*}[t]
 \centering
\includegraphics[width=0.98\textwidth]{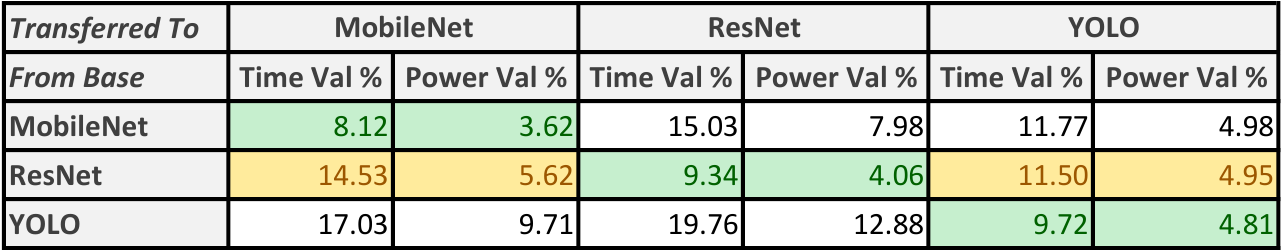}
%\vspace{-0.1in}
\caption{Effectiveness of PowerTrain using different \textit{reference DNN workloads} to transfer from. MAPE for time and power predictions validated against ground truth is reported.}

\label{fig:tltable}
%\vspace{-0.1in}
\end{figure*}

\subsection{Choice of Base DNN Workload for PowerTrain}

The choice of the \textit{reference DNN workload} to train PowerTrain on can influence the quality of transfer learning. We have three candidate DNN workloads, ResNet, MobileNet and YOLO for which we have $4.4k$ training samples to allow reference model training. In Figure~\ref{fig:tltable}, we compare the validation MAPE when using each of these three as a reference and transferring to the other two, against the ground truth observed time and power values. We also report the validation MAPE for the reference model itself without any transfer, i.e., equivalent to NN model with all samples; this is likely to be the best possible prediction model.
This is seen in the diagonal elements (green), which have MAPE between $8.1$--$9.7\%$ for time predictions and $3.6$--$4.8\%$ for power. 

Among the off-diagonal elements using PowerTrain transfer learning, \resnet proves to be the best candidate as reference model. It results in time and power MAPE of $14.5\%$ and $5.6\%$ for \mobilenet and $11.5\%$ and $4.9\%$ for \yolo (highlighted in yellow). In contrast, picking \mobilenet as the reference results in time and power MAPE of $15.0\%$ and $7.9\%$ for \resnet and $11.8\%$ and $4.9\%$ for \yolo. \yolo performs even worse as a base model. Intuitively, we reason that \resnet has the highest variation in observed power values across the different power modes, and is able to generalize better to other DNNs which operate over a limited power range during training. Unless stated otherwise, in subsequent experiments, PT uses \resnet as the reference DNN.

\begin{figure*}[t]
 \centering%~%

\subfloat[ResNet*]{%
    \includegraphics[width=0.5\textwidth]{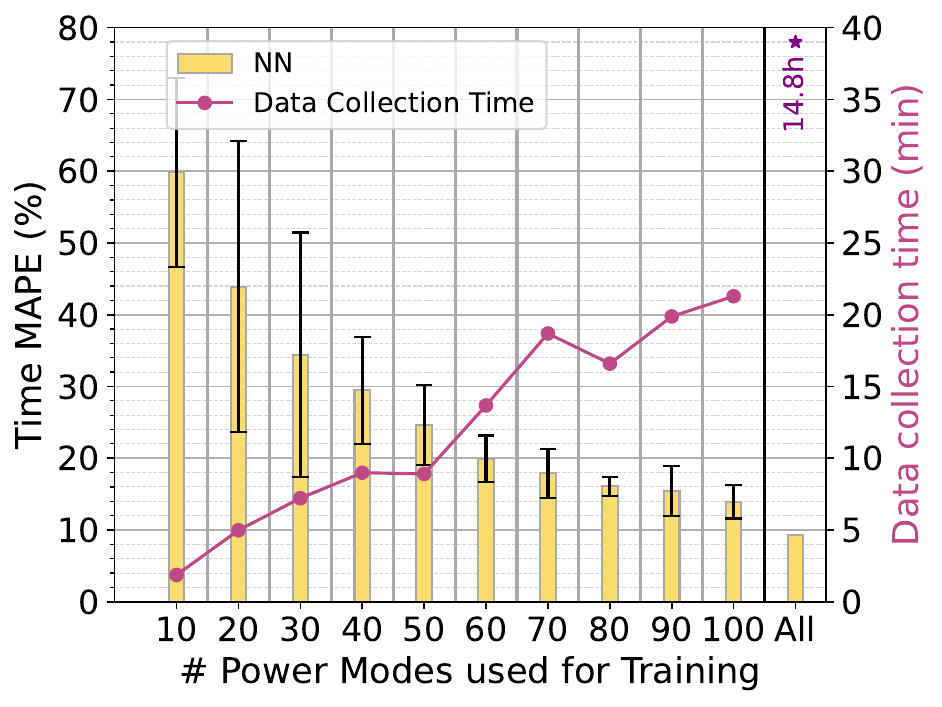}
    \label{fig:ml:resnet:time}
  }%
\subfloat[MobileNet]{%
    \includegraphics[width=0.5\textwidth]{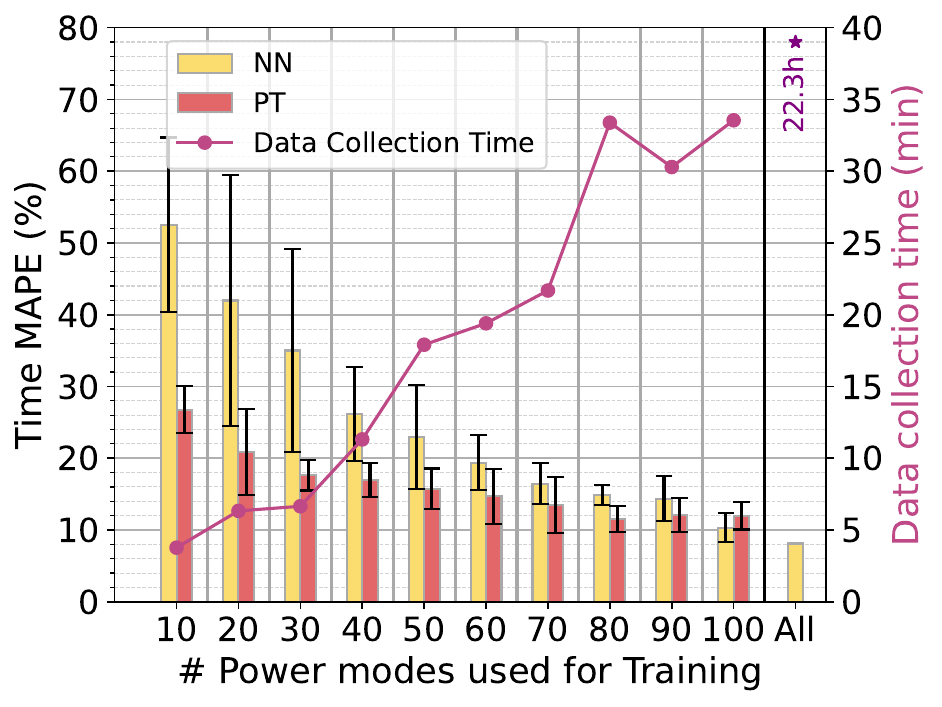}
    \label{fig:ml:mobnet:time}
  }\\
\subfloat[YOLO]{%
    \includegraphics[width=0.49\textwidth]{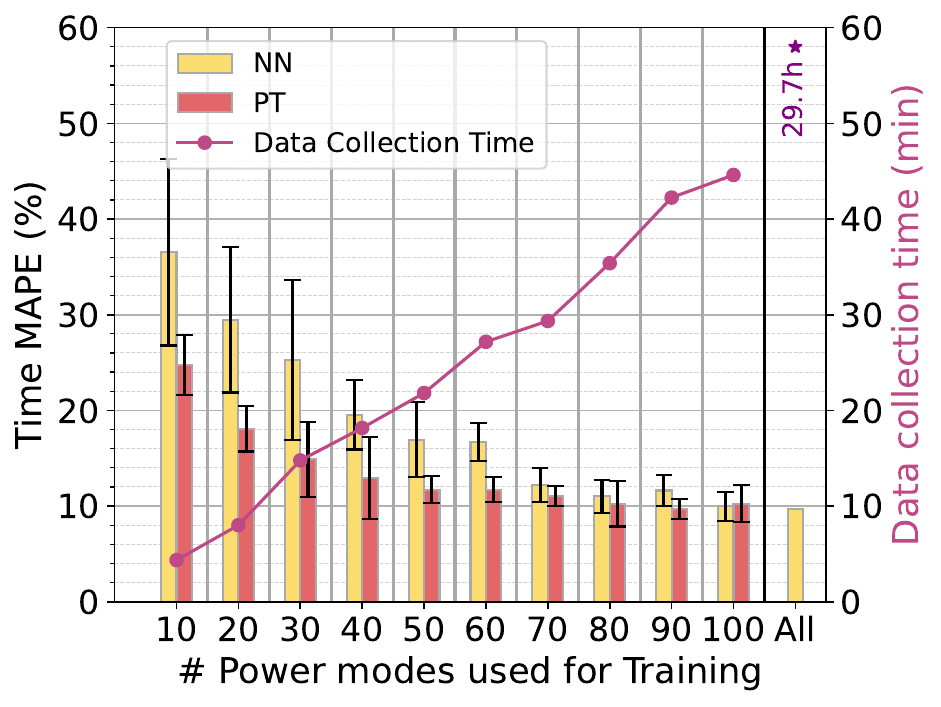}
    \label{fig:ml:yolo:time}
  }%

\caption{Prediction error and profiling overheads for \textit{time prediction} for different DNN workloads}

\label{fig:ml:time}

\end{figure*}

\begin{figure*}[t]
 \centering%~%
\subfloat[ResNet*]{%
    \includegraphics[width=0.5\textwidth]{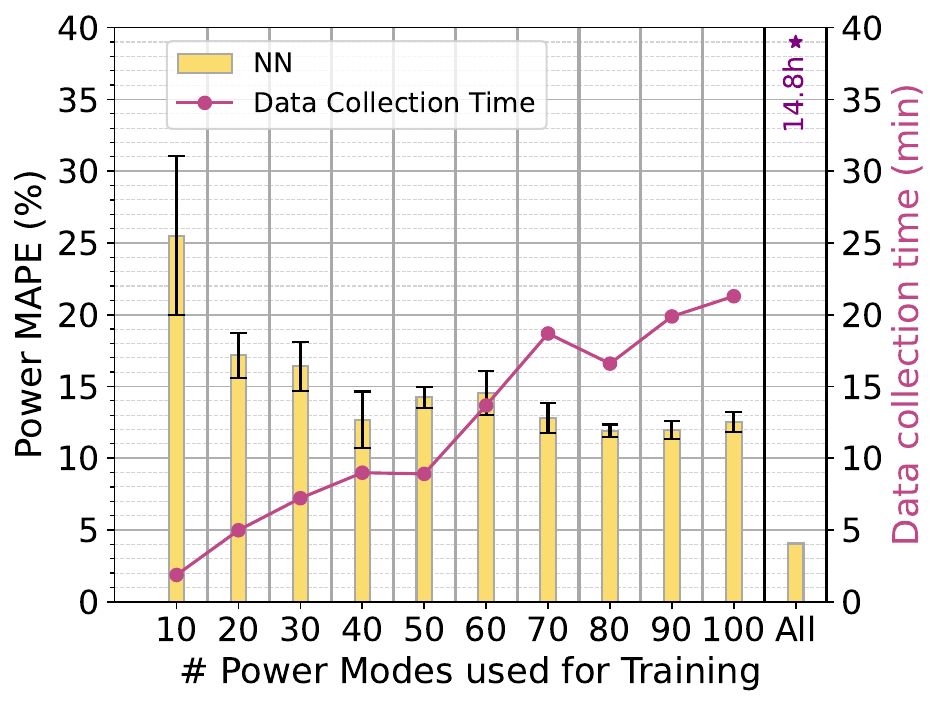}
    \label{fig:ml:resnet:power}
  }%
\subfloat[MobileNet]{%
    \includegraphics[width=0.5\textwidth]{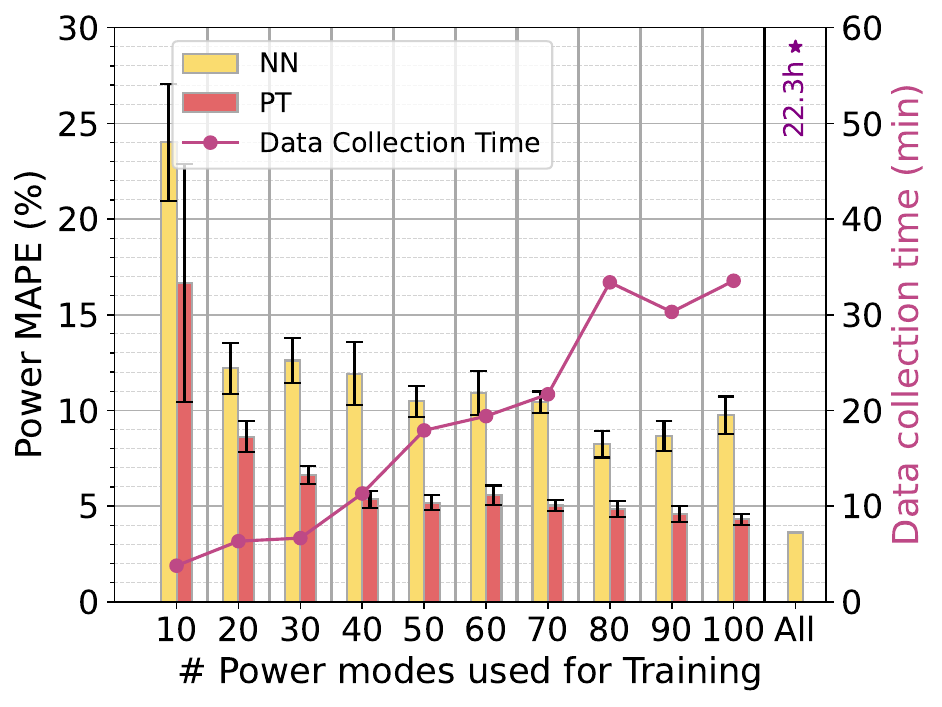}
    \label{fig:ml:mobnet:power}
  }\\
% Need to update to yolo image
\subfloat[YOLO]{%
    \includegraphics[width=0.50\textwidth]{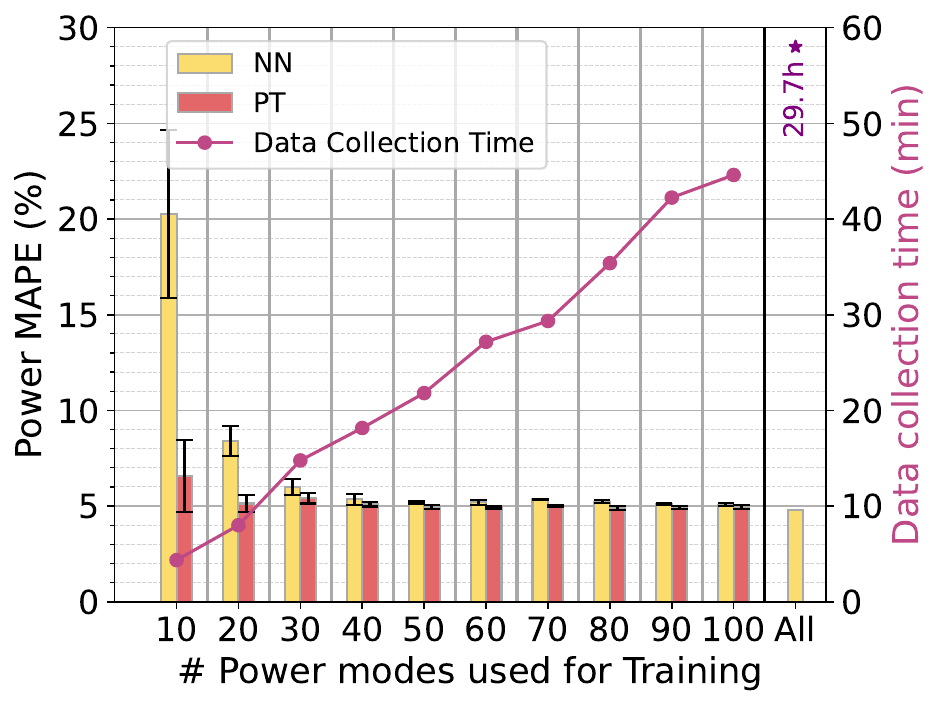}
    \label{fig:ml:yolo:power}
  }
%\vspace{-0.1in}
\caption{Prediction error and profiling overheads for \textit{power prediction} for different DNN workloads}

\label{fig:ml:power}
%\vspace{-0.1in}
\end{figure*}

\subsection{Time and Power Prediction Accuracy vs. Profiling Overheads}
Here, we compare the performance of NN model trained using \textit{All} profiling samples and with different smaller samples for the three default DNN workloads, against PowerTrain (PT) trained on ResNet as reference and transferred to the other two, with a different number of training samples used for transfer learning. Figures.~\ref{fig:ml:time} and~\ref{fig:ml:power} report the median MAPE for validation of the prediction models on all $4386$ power modes across $10$ training and validation runs. 
Here, the X axis indicates the size of the training data used to train the model (NN), or update the transferred model (PT) from the base \resnet NN mode. \textit{All} indidates NN training on all $4.4k$ samples. The median MAPE is shown as bars on the left Y axis with whiskers indicating $Q1$--$Q3$ error ranges, while the profiling time taken for these many training is shown on the right Y axis (line plot) as overheads. In both cases, smaller is better. Since ResNet is the reference model, we do not report PT results when ResNet is the target DNN workload.

\claim{PowerTrain performs better than NN on time predictions for a new DNN with fewer number of power modes profiled.} 
In Figure~\ref{fig:ml:time}, the MAPE for PT with $10$ power modes as training input for \mobilenet is $26.7\%$ while the error for NN using $10$ power modes is $52.6\%$. With just $30$ power modes, PT improves to a MAPE of $<20\%$ for \mobilenet while NN has a much higher $35\%$ error. Similarly, for \yolo, at $30$ power modes, PT's MAPE is $<15\%$, while at $10$ samples it is $24.7\%$. Additionally, the $Q1$--$Q3$ whisker is much tighter for PT than NN, showing less variability on the accuracy even with different random subsets of power modes picked for training. This shows the potential of PowerTrain to generalize from a model trained fully on one reference DNN to another with few samples. The data collection time (right Y axis, line) for $30$ power modes is under $15$~mins for all $3$ DNNs, and under $25$~mins for $50$ power modes. A typical training to convergence runs for $100$s of epochs and many hours. Hence, the data collection overhead is a smaller fraction. 

\claim{PowerTrain for power predictions achieves a higher accuracy for a new workload with fewer profiling samples than NN.}
In Figure~\ref{fig:ml:mobnet:power}, $20$ power mode samples for \mobilenet allows PowerTrain to have a MAPE of just $8.5\%$, as compared to $12\%$ for NN trained on a similar number of samples. Similarly, Figure~\ref{fig:ml:yolo:power} for \yolo shows PT achieve a power MAPE of $6.8\%$ with 10 samples compared to $21\%$ for NN. At larger number of samples of 50 or beyond, they have similar accuracies for YOLO though PT maintains a consistent $5\%$ improvement over NN for MobileNet.

\claim{PT power prediction has lower MAPEs ($5$--$10\%$) as compared to time predictions ($10$--$20$\%) across all DNNs.}

From Figures~\ref{fig:ml:mobnet:power} and~\ref{fig:ml:mobnet:time}, at $10$ power mode samples for \mobilenet, we observe a $2\times$ lower MAPE for power as compared to time. This holds across DNN workloads, and for both PT and NN. At $50$ power modes, which will be our default configuration for PT going forward, the power MAPEs for \mobilenet and \yolo are $5.2\%$ and $4.9\%$ in contrast to time MAPEs of $15.7\%$ and $11.7\%$ for \mobilenet and \yolo, respectively. This can be explained by the fact that we see little variation in power values for a given power mode, enabling the ML-based techniques to predict more accurately.

\claim{By $100$ power modes, PowerTrain reaches close to the optimum accuracy for time and power prediction, and is comparable to NN training on all $4.4k$ power modes profiled.}
In Figure~\ref{fig:ml:mobnet:time}, at $100$ power modes, PT has a MAPE of just $11.9\%$ while NN trained on all samples has a MAPE of $8.1\%$. For power predictions on \mobilenet, PT at $100$ samples has a MAPE of $4.3\%$ while the NN on all samples has a MAPE of $3.6\%$. For \yolo, the MAPE for time predictions is $10.18\%$ for PT at $100$ samples while the NN on all is a close $9.7\%$, and a similar trend is seen for power prediction as well. This again demonstrates the ability of PT to give higher accuracy with fewer samples.

\subsection{Generalization of PowerTrain}

A vital benefit of PowerTrain is its ability to generalize to heterogeneous DNN workloads using a few profiling samples to achieve a higher accuracy. We next evaluate this generalizability along three dimensions (Figure~\ref{fig:workflow:TLmnemonic}), each being more diverse compared to the reference DNN workload trained: (1) \textit{Either} the DNN architecture or the dataset being different from the reference DNN workload, (2) \textit{Both} the DNN architecture and dataset being different, and (3) The \textit{edge device} and the \textit{DNN workloads} being different. The latter is particularly beneficial since it can encompass a wide range of DNN workloads and devices, e.g., federated learning over heterogeneous edge accelerators.

\begin{figure*}[t!]
%\vspace{-0.1in}
\centering
\subfloat[Overlapping Workloads]{%
    \quad\includegraphics[width=0.45\textwidth]{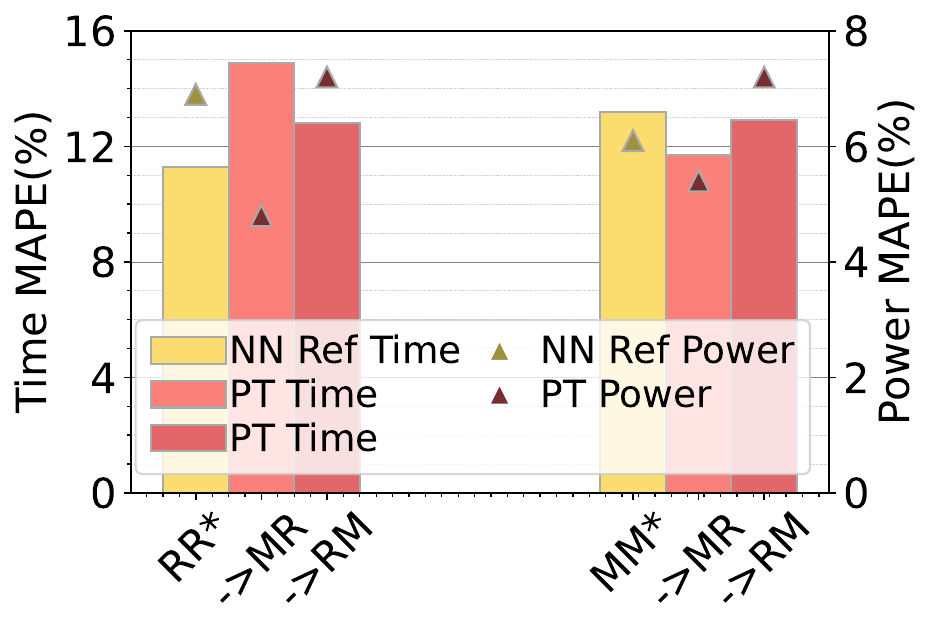}\quad
    \label{fig:gen:wl}
  }\hfill
\subfloat[Unseen Workloads]{%
    \quad\includegraphics[width=0.4\textwidth]{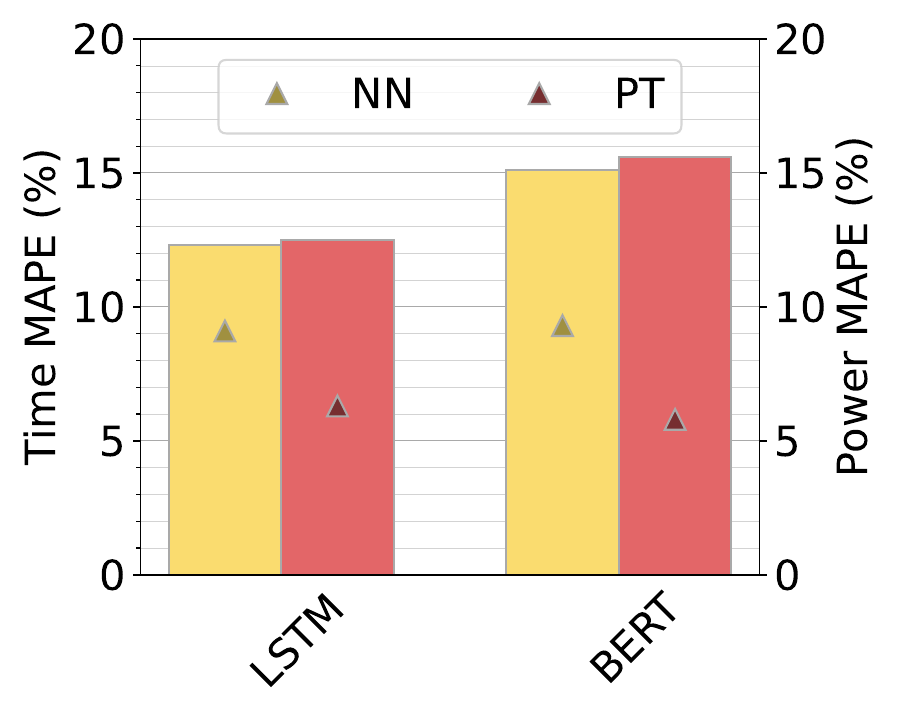}\quad
    \label{fig:gen:dnn}
  }\\
  \subfloat[\addc{Workloads with different training minibatch sizes}]{%
    \quad\includegraphics[width=0.4\textwidth]{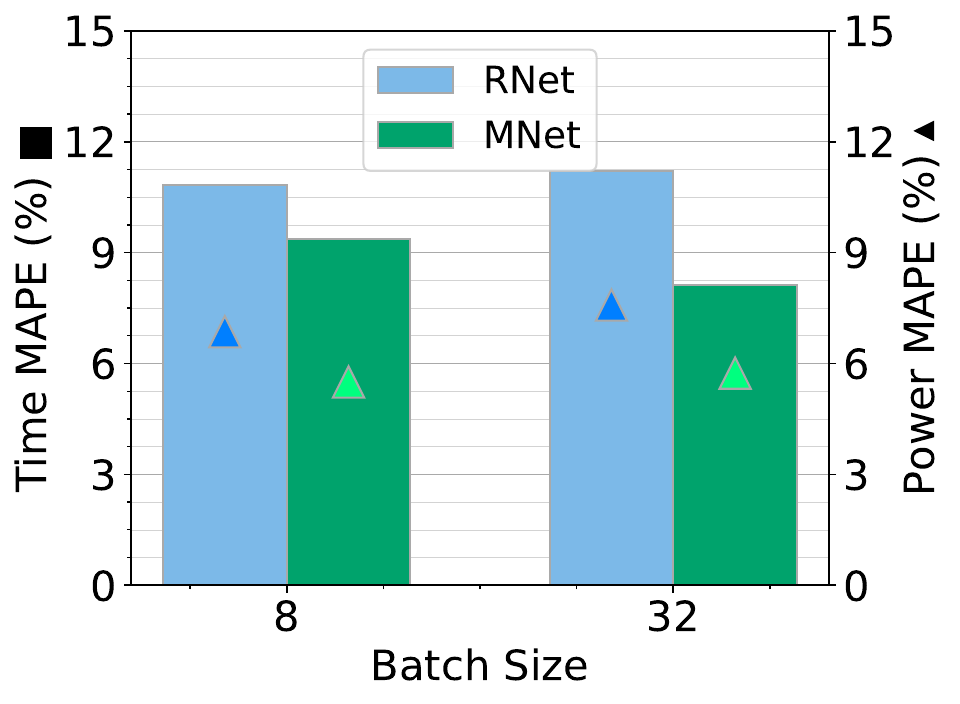}\quad
    \label{fig:gen:bs}
  }\\
\subfloat[Inter-device, Orin to Xavier]{%
    \quad\includegraphics[width=0.40\textwidth]{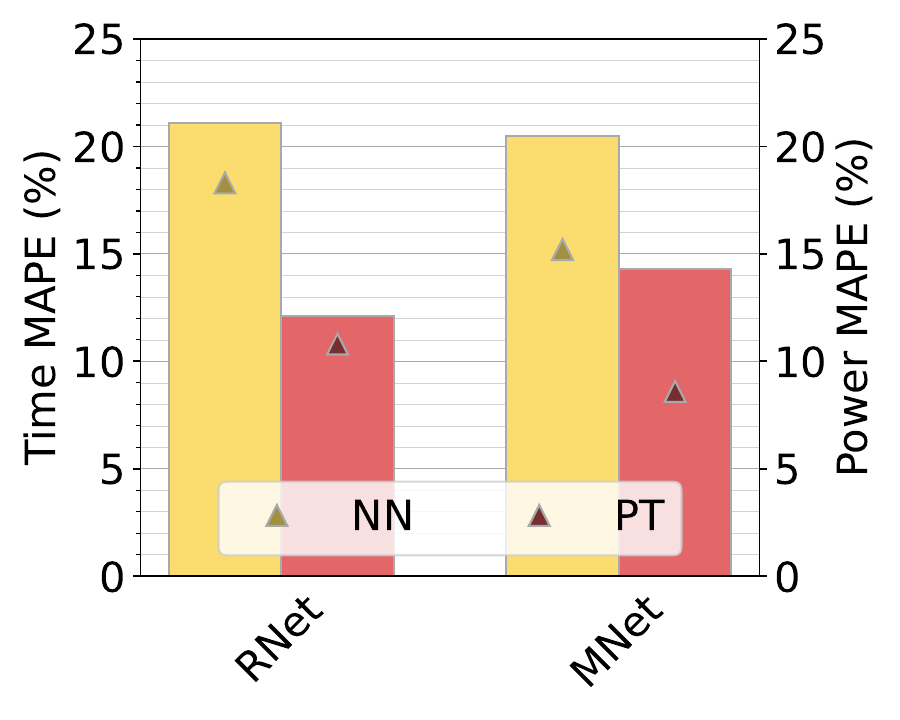}\quad
    \label{fig:gen:dev}
  }
  \vspace{0.15cm}
\subfloat[\addc{Inter-device, Orin to Nano}]{%
    \quad\includegraphics[width=0.40\textwidth]{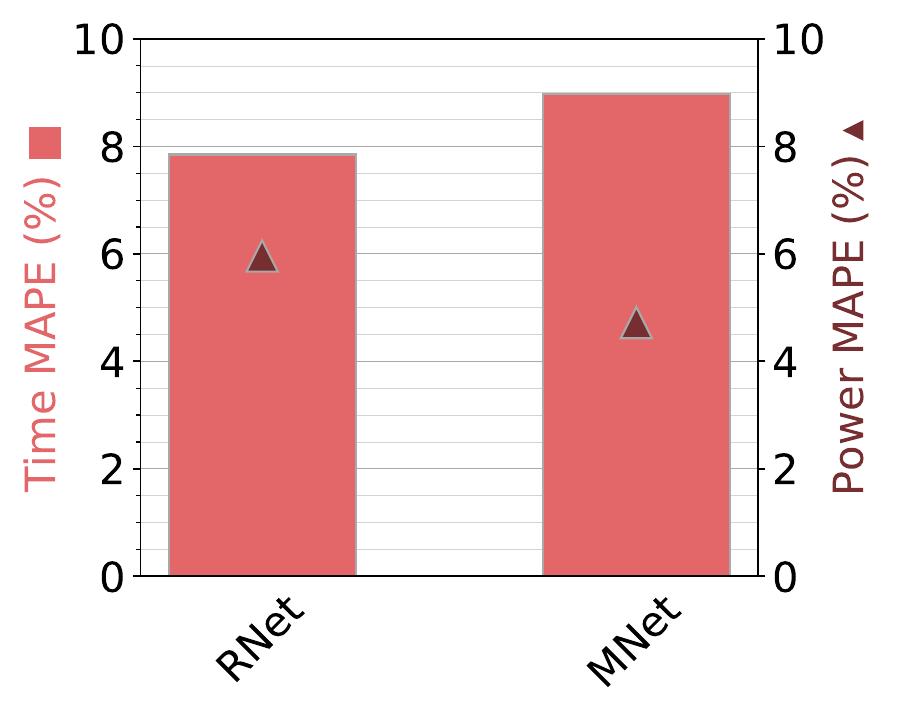}\quad
    \label{fig:gen:nano}
  }\quad

\caption{\modc{PowerTrain generalization when reference model is transferred to new workloads, devices and minibatch sizes.}}

\label{fig:gen:dnndev}

\end{figure*}

\subsubsection{Different DNN Architecture or Dataset from Reference}

Here, PowerTrain uses two different reference NN models: one trained on ResNet with its native dataset ImageNet (Table~\ref{tbl:modeldataset}), which we term as $RR*$, and another trained on MobileNet with its native GLD dataset ($MM*$). Now, PT tries to use each of these reference models and transfer to a new workload where either the DNN architecture or the dataset is different~\footnote{Both ImageNet and GLD are image datasets that are suitable for training classifiers, and can be used with both ResNet and MobileNet architectures.}. These combinations are listed as $RR* \rightarrow RM$, when transferring to ResNet architecture with MobileNet's native dataset GLD, $RR* \rightarrow MR$ when transferring to MobileNet architecture with ResNet's native dataset ImageNet, and similarly $MM* \rightarrow MR$ and $MM* \rightarrow RM$. So, either the architecture or dataset overlaps with the reference. We use 50 random power mode samples for transferring using PT. We report the time and power MAPE for these scenarios when validated against 100 samples of power modes, and also offer the MAPE for the reference DNN $RR*$ and $MM*$, which serves as the best case.

\claim{PowerTrain generalizes equally well if either DNN or the dataset changes.}
We report the time MAPEs on the left Y axis as bars and the Power MAPEs on the right Y axis as markers. As seen in Figure~\ref{fig:gen:wl}, having overlapping DNN architectures, $RR* \rightarrow RM$ and $MM* \rightarrow MR$, exhibit a MAPE for time that is not much worse than the reference models, $RR*$ and $MM*$, increasing in error from $11.3$ to $12.8\%$ for the former and in fact improving from $13.2$ to $11.7\%$ for the latter. For the power predictions, there is less than $1\%$ change in MAPE.
Similarly, when the datasets overlap but the DNN architecture changes, i.e., $RR* \rightarrow MR$ and $MM* \rightarrow RM$, we see the time MAPE increases from $11.3\%$ to $14.9\%$ and $13.2\%$ to $12.9\%$, respectively. Here too, the power MAPE does not deteriorate by more than $1\%$.
So PowerTrain generalizes well when either the DNN architecture or the dataset overlaps with the reference.

\subsubsection{Unseen and Diverse DNNs Workloads}
The earlier results in Figures~\ref{fig:ml:time} and~\ref{fig:ml:power} showed PT transfer from ResNet reference to MobileNet and YOLO workloads that did not have any workload overlaps. Here, we expand that pool to include two workloads with very different DNN architectures compared to ResNet's CNN base: a BERT transformer architecture and an LSTM RNN-based architecture (Table~\ref{tbl:modeldataset}), and text-based training datasets rather than images.
BERT (Bidirectional Encoder Representations from Transformers) Base Uncased~\cite{bert} is a type of transformer-based DNN designed for natural language processing. It captures contextual information and bidirectional dependencies in text. In our workload, BERT is used for Question Answering, using the SQuAD (Stanford Question Answering Dataset) V2.0 Train Set~\cite{squad}. LSTM (Long Short-Term Memory) is a Recurrent Neural Network (RNN) that is designed for learning temporal patterns over lengthy sequences using memory cells and gating mechanisms. Our workload trains an LSTM model for next-word prediction using training data from WikiText~\cite{wikitext}.

PT uses $50$ randomly profiled power mode samples from BERT and LSTM to transfer from ResNet reference, and we validate the transferred models on $50$ other randomly sampled power modes to report the MAPE for time and power relative to the ground truth. This transfer and validation is repeated $20$ times, and the results reported in Figure~\ref{fig:gen:dnn} in comparison with an NN model as baseline trained using the same $50$ power mode samples.

\claim{PowerTrain generalizes well to diverse DNN workloads with few samples, and outperforms NN on power predictions.}
As seen from Figure~\ref{fig:gen:dnn}, for \lstm, the time MAPE is $12.5\%$ for PT and $12.3\%$ for NN, which are comparable. Similarly, for \bert, the time MAPEs of $15.6\%$ and $15.1\%$ are close.
The benefits of PT is visible for power predictions, where PT has a MAPE of $6.3\%$ while NN is $9.1\%$ for \lstm, and likewise, PT has a $3.5\%$ better MAPE for \bert.

\subsubsection{Unseen Device from a Different Generation}

In this experiment, we use PowerTrain to transfer a reference workload trained on Orin AGX to workloads that run on the Xavier AGX developer kit~\cite{Xavier}, which is the comparable previous generation top-end Jetson device. As shown in Table~\ref{tbl:jetsonspecs}, compared to the Orin, the Xavier edge accelerator has an 8 core (vs. 12 for Orin) custom Carmel ARM CPU, a previous generation CUDA architecture (Volta rather than Ampere), and with $\frac{1}{4}^{th}$ the number of CUDA cores (512 vs. 2048), and a previous generation RAM (LPDDR4 vs. LPDDR5). We randomly profile $1000$ power modes out of the available $29,000$ and collect power and time data for \resnet and \mobilenet workloads in a similar manner as before. We use the ResNet workload trained on Orin AGX as our reference model for time and power predictions. Like before, PowerTrain then uses $50$ random power modes for ResNet (or MobileNet) DNN workload profiled on Xavier to transfer-learn onto and validates the re-trained time and power prediction models using the remaining $950$ power mode samples for that workload. 

\claim{PT generalizes well to a device from a different generation and outperforms NN-based training.}
Figure~\ref{fig:gen:dev}, we see that when PT transfers from the reference DNN trained using ResNet workload on Orin to the same workload on Xavier (device changes), we see a time prediction MAPE of $12\%$ and power prediction MAPE of $11\%$. This is much better than training an NN on just $50$ power mode samples of the DNN workload on Xavier, which reports a MAPE of $21\%$ for time and $18\%$ on power. 
Similar benefits hold when both the device and the workloads are different for PT, where transferring the prediction models to MobileNet workload on Xavier has time MAPE of $14\%$ and power MAPE of $9\%$, which are once again much better than NN.

\subsubsection{\addc{Unseen Device from the Same Generation}}
\addc{We also perform a limited set of experiments to examine the generalizability of PowerTrain to a less powerful accelerated edge device, \textit{Jetson Orin Nano}~\cite{Orin-nano}, but from the same Jetson generation as the Orin AGX. We randomly sample $180$ of the available $1800$ power modes for Orin Nano and collect profiling data for \resnet and \mobilenet workloads. Using the ResNet workload trained on Orin AGX as the reference model for prediction, we transfer-learn to Orin Nano by retraining using $50$ random power modes for ResNet (or MobileNet). We validate the predictions of the transferred model using all $180$ power modes we profile for the workload. As shown in Figure~\ref{fig:gen:nano}, we observe low median time and power errors, with MAPEs of $7.85\%$ and $5.96\%$ for \resnet, and $8.98\%$ and $4.72\%$ for \mobilenet. This confirms the strong generalizability of PowerTrain to even Jetson accelerators that are $6.9\times$ less powerful than the Orin AGX. However, while transferring to such very different device types, hyperparameter tuning is needed during retraining. E.g., when transfer learning from Orin AGX to Orin Nano, the loss metric was changed from MSE to MAPE to achieve this good accuracy.}

This is a powerful result that suggests that there is sufficient similarity \modc{within and across Jetson generations} and DNN workload behaviors that can be learned during reference workload training, and adapted to a much different execution setting. This opens up opportunities to try even more diverse Jetsons like NX and \addc{older} Nano series\delc{, or even different edge models such as Raspberry Pi}. This is left to future work.

\subsubsection{\addc{Unseen Workload Training Minibatch Sizes}}
\addc{We extend the evaluation of PowerTrain to predict the training time and power consumption when using $3$ different minibatch sizes during training: $8,16$ and $32$, which are commonly used. We test this for $2$ DNN workloads, \resnet and \mobilenet. Each new minibatch size is seen as a new DNN training workload.}

\addc{In the first experiment, we use our reference NN, trained on \resnet with a minibatch size of $16$ (\resnet/$16$), and transfer learn onto the same \resnet DNN, but using minibatch sizes of $8$ and $32$ (\resnet/$8$ and \resnet/$32$). We profile and use $50$ power modes for transfer learning. As shown in Figure~\ref{fig:gen:bs}, for \resnet/$8$, our time predictions have a median MAPE of $10.84\%$ and power prediction a median MAPE of $6.86\%$. For \resnet/$32$, the time and power MAPEs are comparable at $11.2\%$ and $7.28\%$, respectively. These are similar to the MAPE for the default \resnet/$16$, which was $8.83\%$ and $7.39\%$ for time and power respectively.}

\addc{In the second experiment, we transfer the same reference NN (\resnet/$16$) to a different DNN workload and with different minibatch sizes, \mobilenet/$8$, /$16$ and~/$32$. PT's predictions generalize well here too. The median MAPE for time predictions range narrowly from $7$--$9.4\%$ and for power between $5.5$--$5.7\%$. This shows that PowerTrain is robust to changes in training minibatch sizes.}

\section{Optimizing DNN Workloads using our Prediction Models} 
\label{sec:results:optimize}

\begin{figure*}[t!]

 \centering%~%

\subfloat[ResNet Ref. Workload on All]{
    \includegraphics[width=0.45\textwidth]{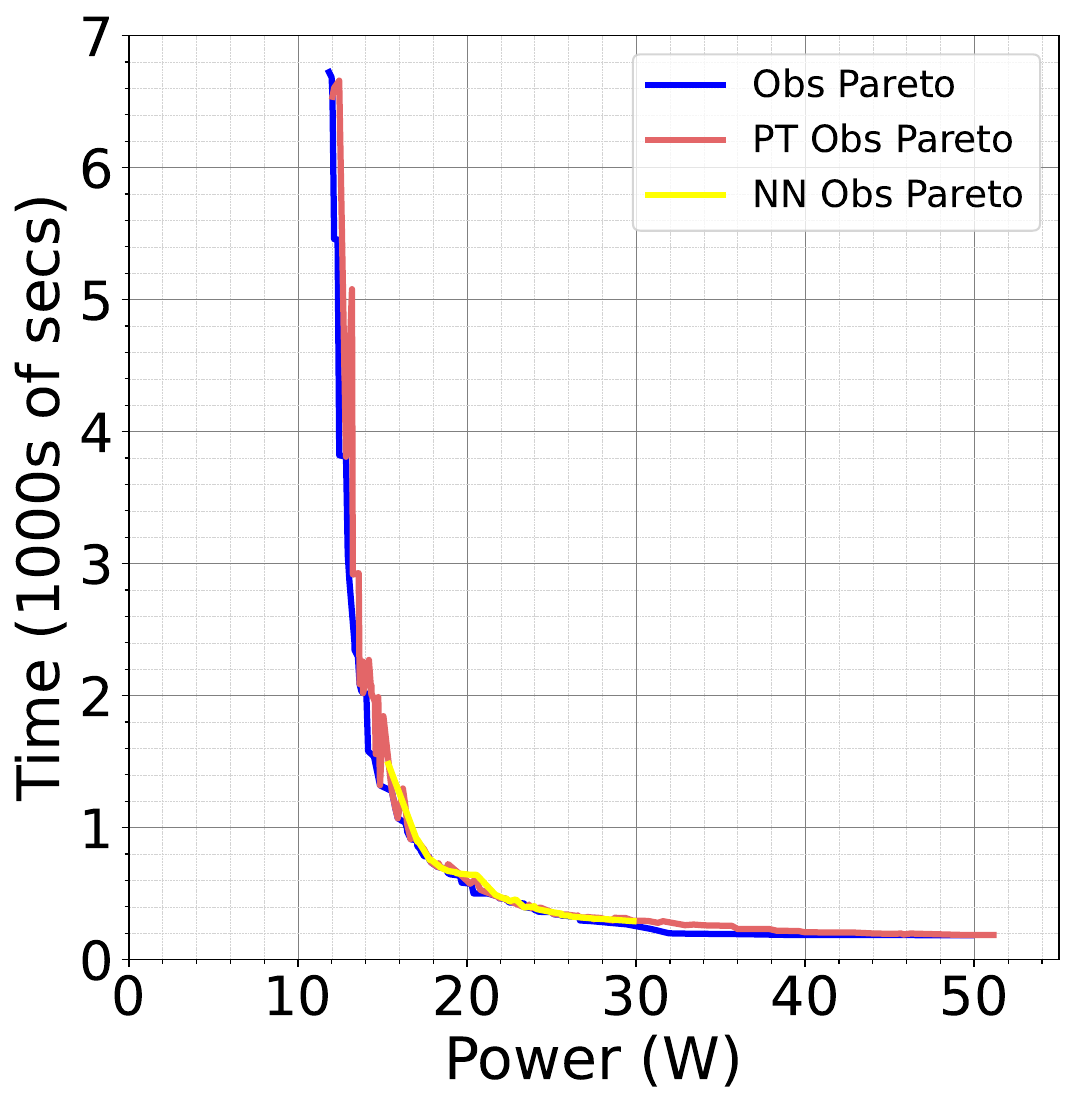}
    \label{fig:pareto:resnet}
  }\qquad
\subfloat[PT onto MobileNet]{
    \includegraphics[width=0.45\textwidth]{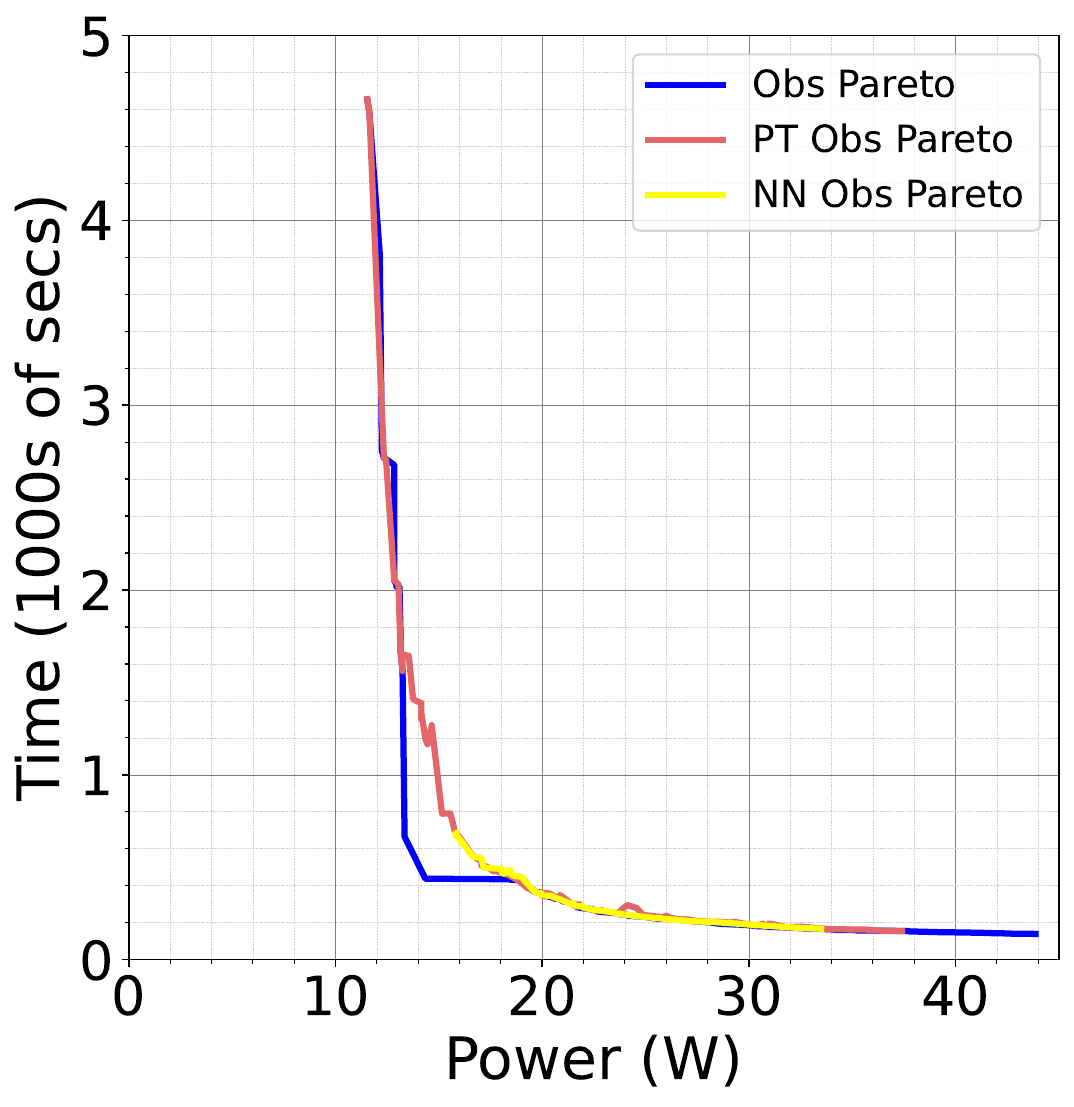}
    \label{fig:pareto:mobnet}
  }\\
  \subfloat[PT onto YOLO]{
    \includegraphics[width=0.45\textwidth]{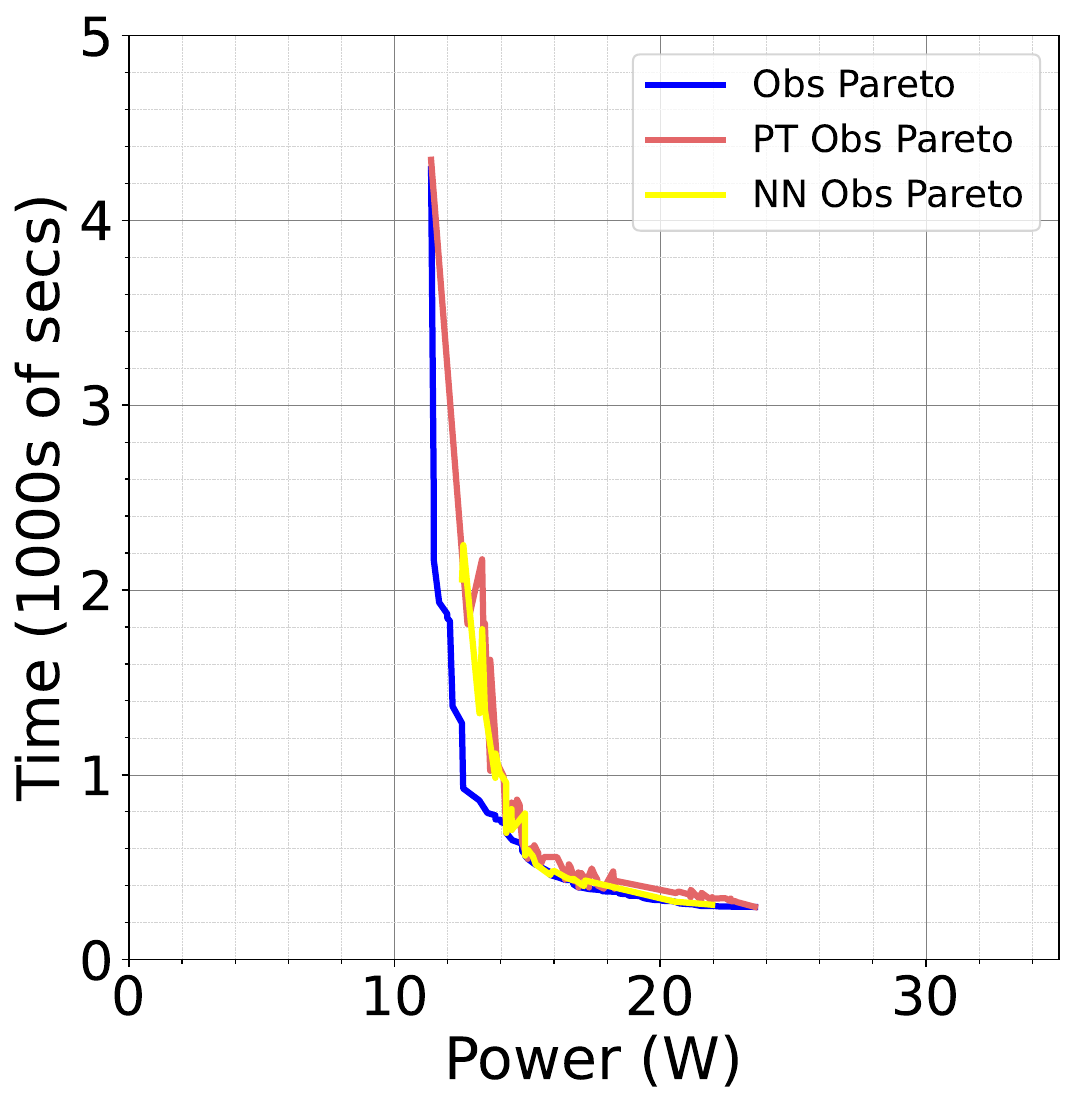}
    \label{fig:pareto:yolo}
  }

\caption{Power--Time Pareto front over scatter plot of predicted training time and predicted power using all power modes, using Transfer Learned model from ResNet NN}

\label{fig:tl:powerpareto}
 %\vspace{-0.15in}
\end{figure*}

\begin{figure}

 \centering%~%
    \includegraphics[width=0.5\columnwidth]{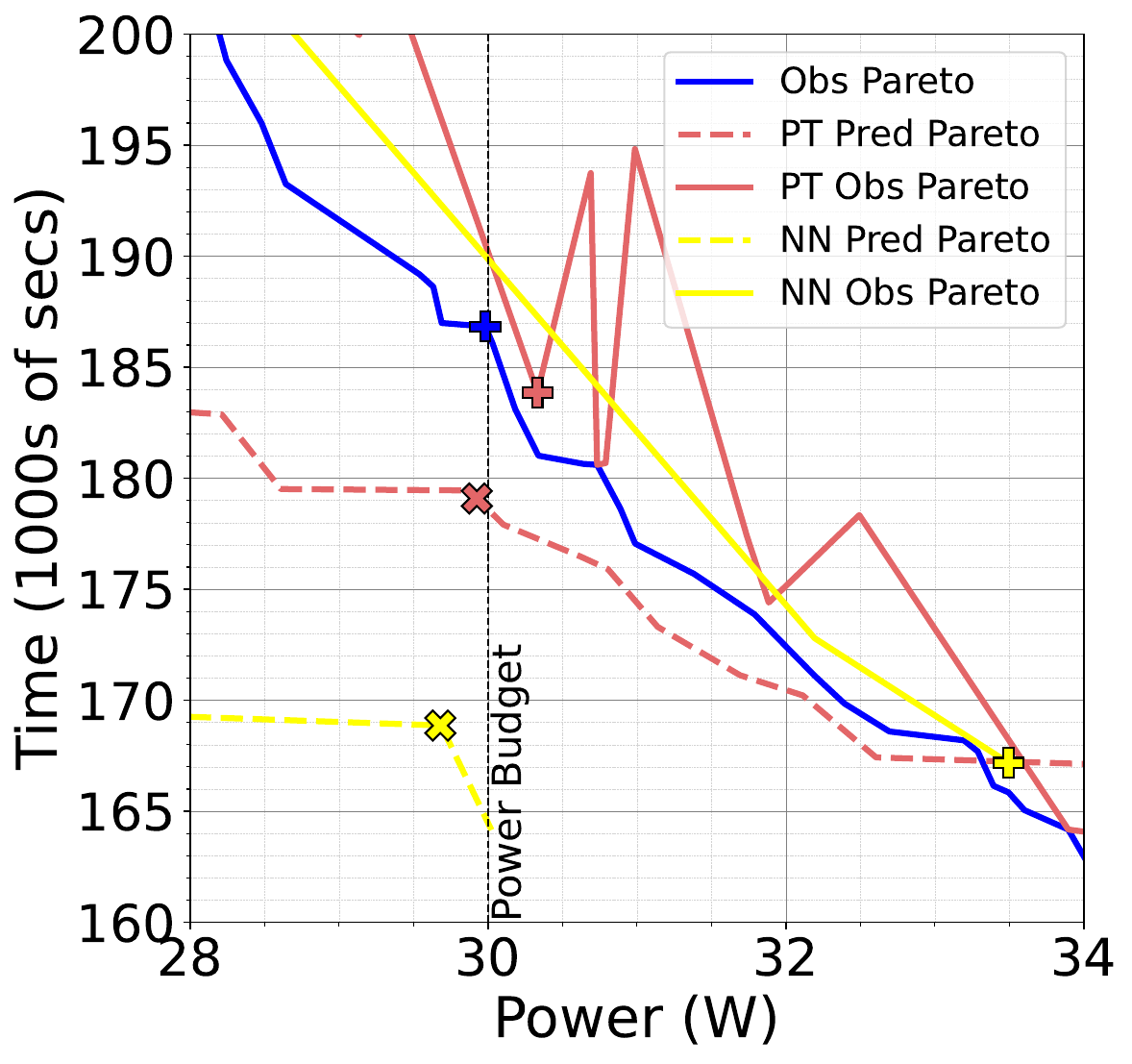}
  
\caption{Zoomed in Power--Time Pareto front for MobileNet for a power target of 30W illustrating the performance of NN and PT wrt ground truth}
\label{fig:paretoex}
\end{figure}

One of the applications of our time and power prediction models is to optimize the power mode configuration of DNN workloads. We formulate an optimization problem as follows. Given a DNN workload $m_{tr}$, choose a power mode $pm$ from a set $\mathbb{PM}$ that minimizes training time per epoch $t_{tr}$ while ensuring that the power load $P_{tr}$ falls within a user-specified budget $P_{b}$, i.e.,
\begin{equation*}\label{eq:train}
    \left.\begin{aligned}
      Given \ & m_{tr}, \\
      \sel \ & pm \in \mathbb{PM}\\
      \min \ \ & t_{tr} \\
      \subto \quad & P_{tr} \leq P_{b}
    \end{aligned} \right.
\end{equation*}

As illustrated in the workflow in Figure~\ref{fig:workflow}, we use our prediction models to solve the optimization problem. For PowerTrain, we use the reference model trained using the ResNet workload on $4368$ power modes, and transfer to other target workloads using $50$ random power mode profiling samples. Since the time predictions are per minibatch, we scale them to per epoch times. As a ground truth optimization solution, we draw the observed Pareto front (\textit{Obs Pareto} in Figure~\ref{fig:tl:powerpareto}) using the profiling information for $4.4k$ power modes to minimize time and power. These set of points offer the least time value for a given power, and vice versa. For various power limits specified by the user, we can trivially search Pareto points to find the \textit{optimal power mode} with a power value that is closest to but less the power budget and report the training time for this.

Using the prediction models, we follow a similar approach, with the key difference that the models are used to estimate the training time and power for all possible ($4.4k$) power modes. We build a similar Pareto from PT predictions and this predicted Pareto (
\textit{PT Pred Pareto} in Figure~\ref{fig:paretoex} \addc{and 3(c) in Figure~\ref{fig:workflow}}) is used for the optimization decisions. \addc{Specifically, for a user-specified power limit (3(a)), we perform a lookup on the Predicted Pareto to find the optimal power mode (2(d)).}. We also report the matching ground-truth values for these predicted power modes on the Pareto, shown as \textit{PT Obs Pareto} in Figures~\ref{fig:tl:powerpareto} and \ref{fig:paretoex}. The NN Prediction model using 50 samples serves as a baseline to compare against, and is shown as \textit{NN Obs Pareto} and \textit{NN Pred Pareto}.

\subsection{Baselines}

In addition, we also offer a few baseline approaches for solving the optimization problem to contrast against PowerTrain. \textbf{MAXN} is the default power mode setting on the Jetson Orin AGX, and offers the best time performance (albeit with a high power load).
\textbf{Random Sampling Pareto (RND)} is a simple sampling-based baseline where we randomly profile $50$ power modes and build a Pareto front from these, which is then used for the optimization.
We continue to use the \textbf{NN model} custom trained for each DNN workload using $50$ random power modes, and use this as a baseline to predict the Pareto front over all power modes and perform optimization.

\subsection{Results}
We solve the optimization problem for a parameter sweep of power limits, from $17$W to $50$W in increments of $1$W, to get a set of optimal solutions for the set of problems, each having a different power limit. We compare the results of the optimization solutions using our approaches and the baselines against the optimal solution given by the ground-truth Pareto. We report several metrics: the \textit{time penalty\%}, which is the excess training time spent for the power mode selected by a strategy compared to the optimal power mode; the \textit{normalized area under the curve (AUC)} of power in excess of the given budget for the given solution by a strategy (W per solution); the \textit{\% of solutions that exceeds the power limit} for a strategy (A/L); and the \textit{\% of solutions that exceed the limit by more than a 1-Watt threshold} (A/L+1). There are no power errors for the observation-based (rather than prediction based) random sampling.

\begin{comment}

\end{comment}

\begin{figure*}[t]
 \centering%~%

\subfloat[\resnet]{%
    \includegraphics[width=0.25\textwidth]{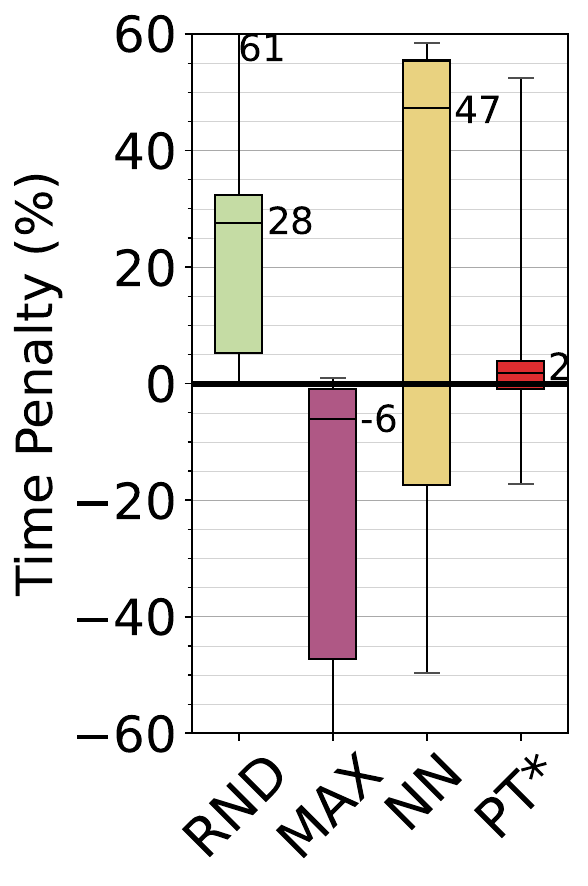}
    \label{fig:paretoerr:resnet_time}
  }~
 \subfloat[\mobilenet]{%
    \includegraphics[width=0.25\textwidth]{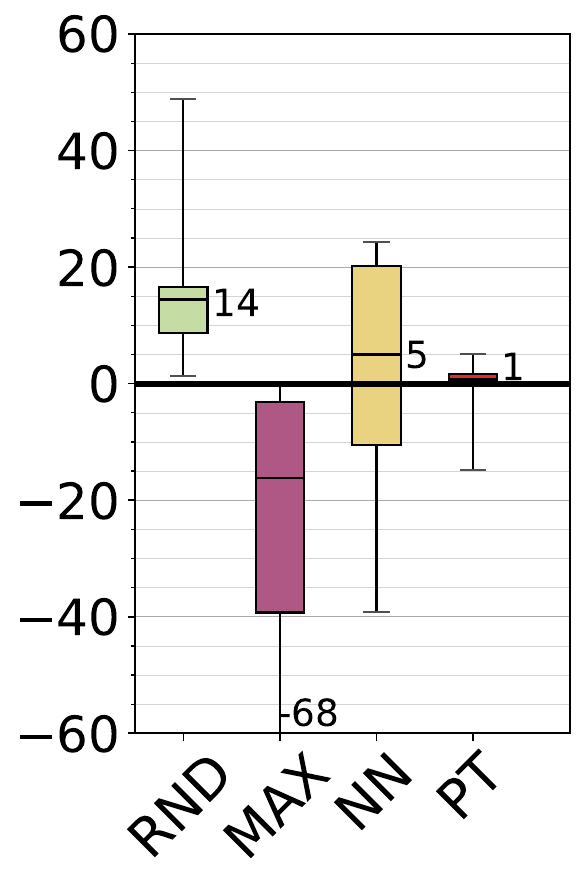}
    \label{fig:paretoerr:mobnet_time}
  }~
  \subfloat[\yolo]{%
    \includegraphics[width=0.25\textwidth]{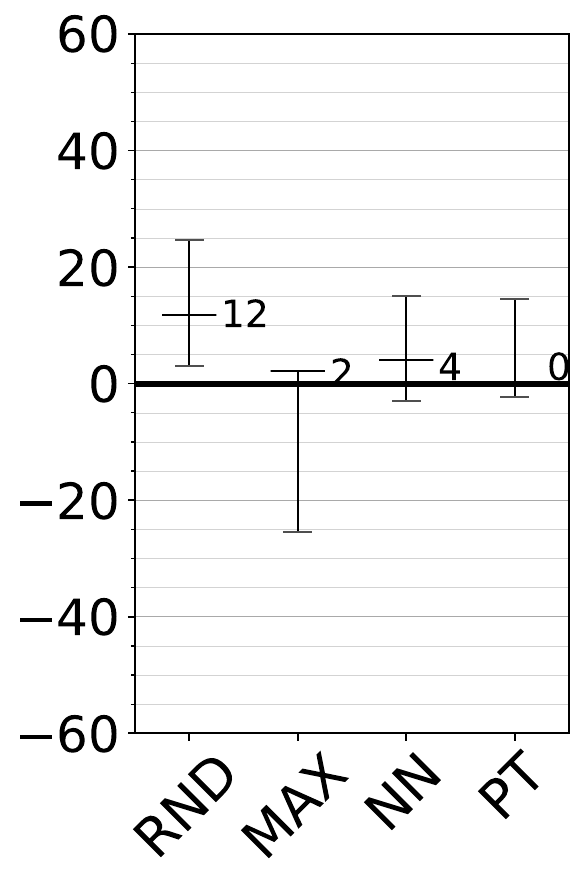}
    \label{fig:paretoerr:yolo_time}
  }\\
\subfloat[\lstm]{%
    \includegraphics[width=0.20\textwidth]{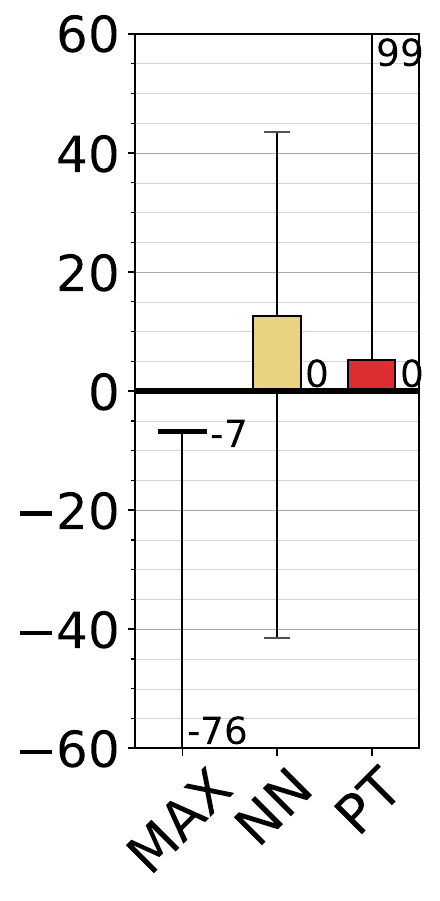}
    \label{fig:paretoerr:lstm_time}
  }
  \subfloat[\bert]{%
    \includegraphics[width=0.20\textwidth]{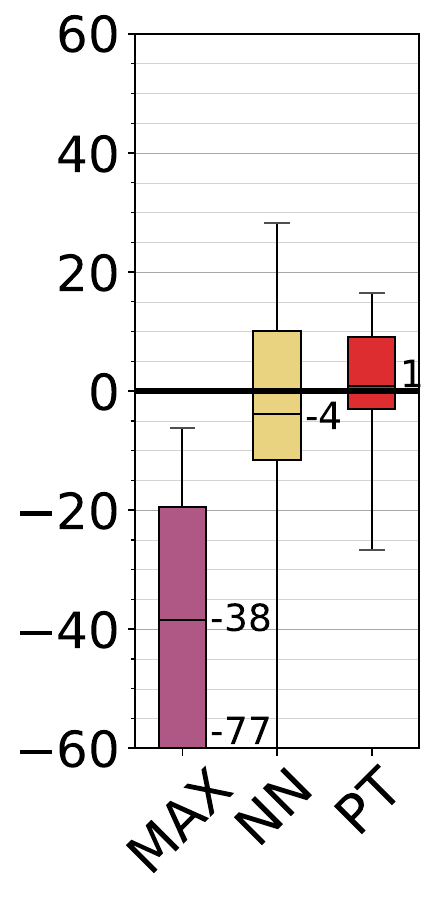}
    \label{fig:paretoerr:bert_time}
    }~
   \subfloat[RNet+MNet]{%
    \quad\includegraphics[width=0.20\textwidth]{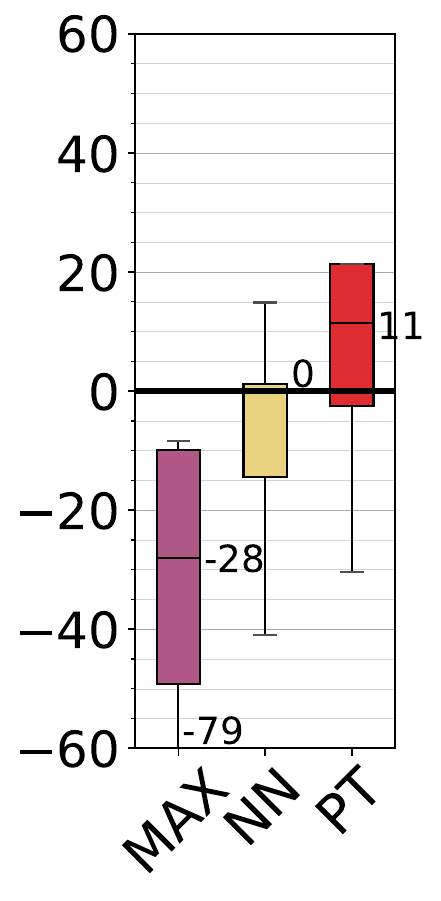}\quad
    \label{fig:paretoerr:resnetgld_time}
  }~
  \subfloat[MNet+RNet]{%
    \quad\includegraphics[width=0.20\textwidth]{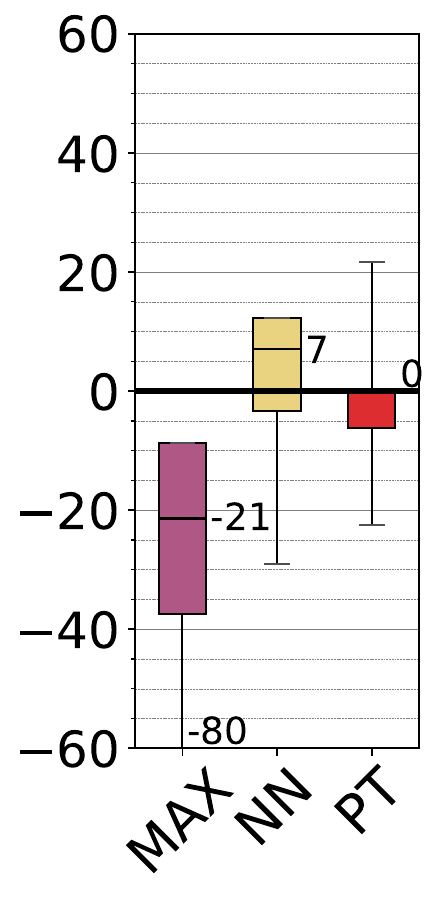}\quad
    \label{fig:paretoerr:mobnetin_time}
  }

\caption{\textit{Time Penalty \%} for model-based optimization relative to ideal Pareto time. Lower is better. *PT for ResNet indicates training of base model on full data.}

\label{fig:pareto:boxtime}

\end{figure*}

\begin{figure*}[t]
 \centering%~%
\subfloat[\resnet ]{
    \includegraphics[width=0.28\textwidth]{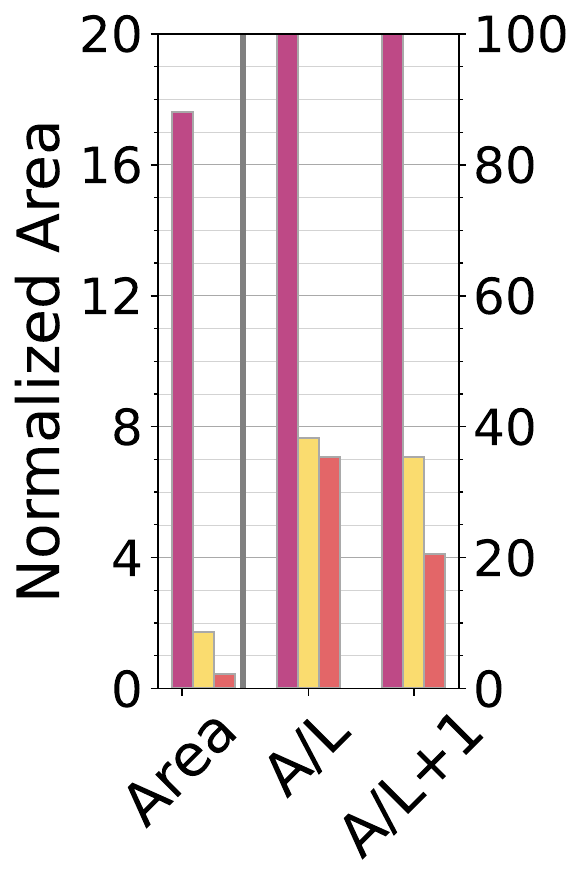}
    \label{fig:paretoerr:resnet_power}
  }~
 \subfloat[\mobilenet ]{
    \includegraphics[width=0.28\textwidth]{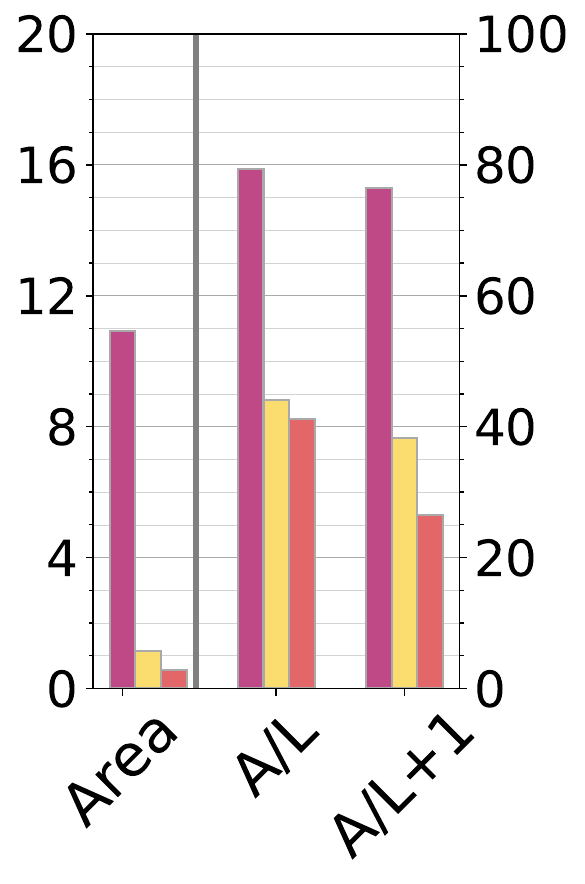}
    \label{fig:paretoerr:mobnet_power}
  }~
  \subfloat[\yolo]{
    \includegraphics[width=0.28\textwidth]{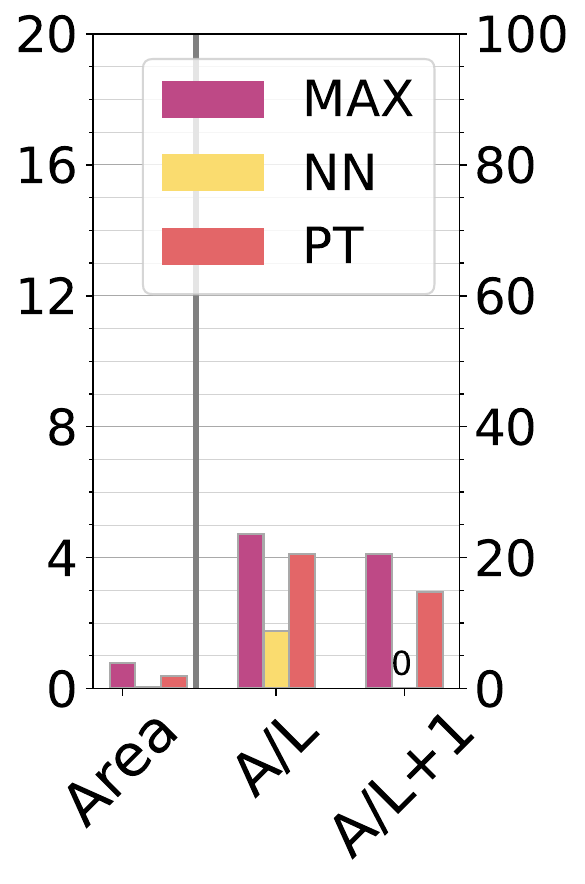}
    \label{fig:paretoerr:yolo_power}
  }\\
\subfloat[\lstm]{
    \includegraphics[width=0.23\textwidth]{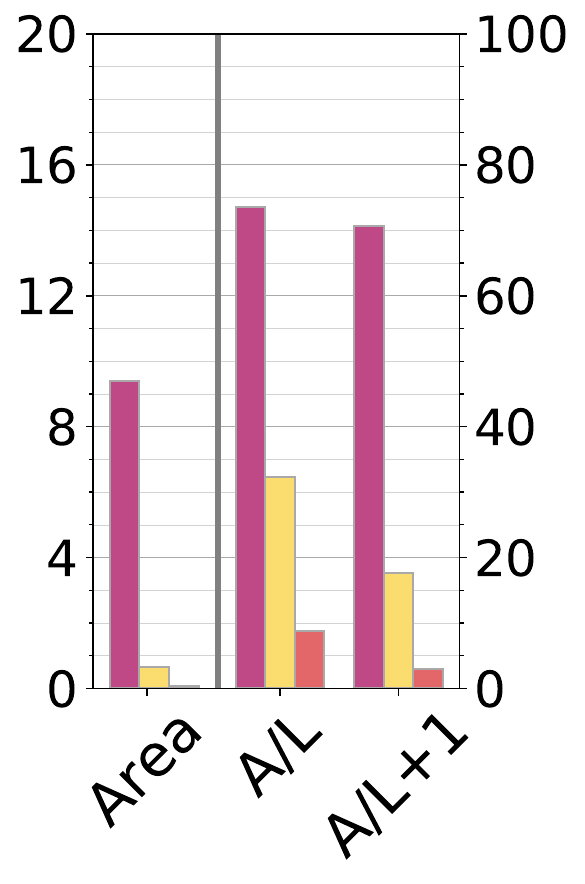}
    \label{fig:paretoerr:lstm_power}
  }~
  \subfloat[\bert]{
    \includegraphics[width=0.23\textwidth]{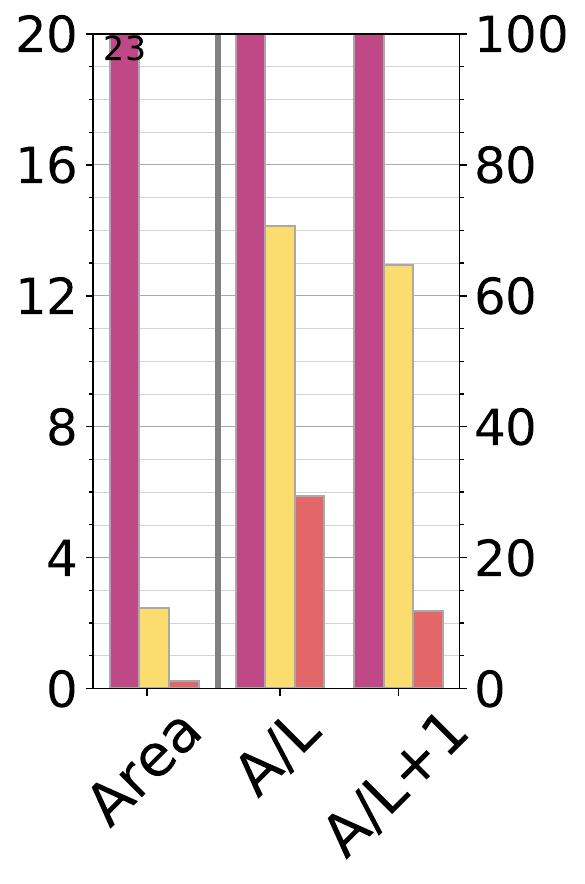}
    \label{fig:paretoerr:bert_power}
    }~
   \subfloat[RNet+MNet]{%\resnet + GLD
    \includegraphics[width=0.23\textwidth]{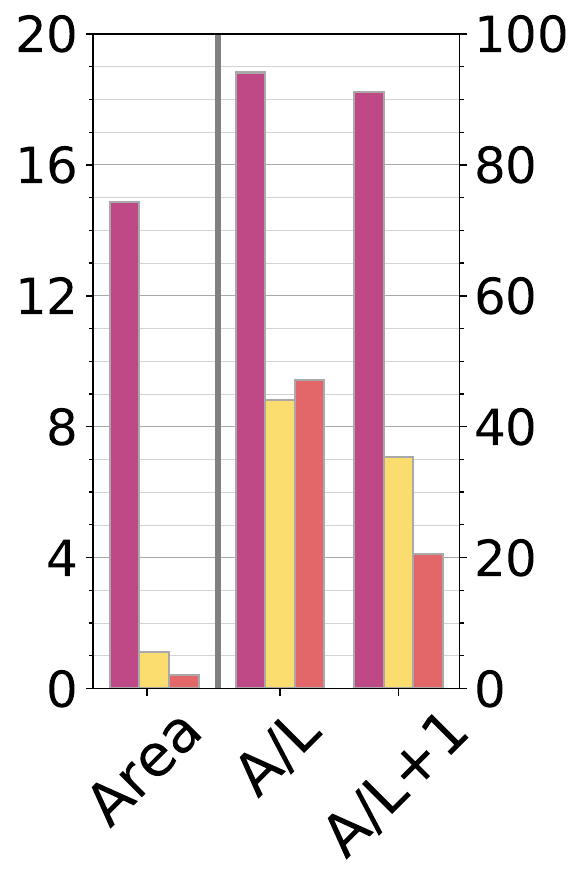}
    \label{fig:paretoerr:resnetgld_power}
  }~
  \subfloat[MNet+RNet]{%\mobilenet + IN
    \includegraphics[width=0.23\textwidth]{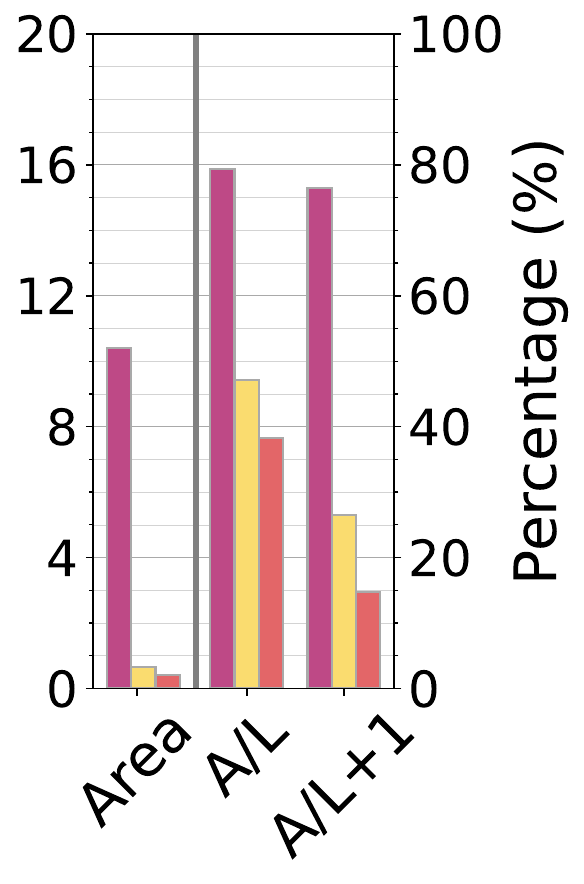}
    \label{fig:paretoerr:mobnetin_power}
    }
\caption{Pareto Power Errors for different DNN workloads} 
\label{fig:pareto:barpower}
\end{figure*}

\claim{PT observed Pareto is close to the ground truth Pareto while it deviates for NN, leading to better optimization results for PT.}
In Figure~\ref{fig:tl:powerpareto}, we draw the \textit{Observed Pareto} using our full profiling information (blue line). Further, we come up with the prediction Paretos for NN and PT and report the matching ground-truth values for these predicted power modes on the Pareto, shown as \textit{NN Obs Pareto} and \textit{PT Obs Pareto}, respectively. As can be seen, PT closely follows the Observed Pareto while NN is limited to a small region.

Further, in Figure~\ref{fig:paretoex}, we present a specific instance of the optimization problem for \mobilenet at a power budget of $30$~W by zooming into the Pareto plot. In this figure, we report the predicted Paretos along with their observed counterparts. The ground truth optimal solution takes $186$~s per epoch and consumes $29.9$~W. The NN optimization chooses a power mode predicted to take $168.9$~s and consume $29.7$~W. However, in reality, this power mode takes a comparable $167.2$~s but consumes $33.5$~W, which overshoots the power budget. The PT optimization chooses a power mode predicted to take $179.1$~s and consume $29.9$~W. In reality, this power mode takes $183.9$~s and consumes $30.3$~W (marginal overshoot by $0.3$~W), which is very close to the optimal.

\claim{PT optimization results in a lower time penalty as compared to NN in most cases.}
As seen from Figure~\ref{fig:pareto:boxtime}, for \mobilenet, the median time penalty for PT is just $0.7\%$ over the optimal as compared to $5\%$ for NN. 
Similarly, for \yolo, PT is as good as optimal with $0\%$ penalty while NN has a median time penalty of $4\%$.
For \lstm and \bert, PT has the same median time penalty as NN, however the distribution is tighter for PT. For \mobilenet using ResNet training dataset (Figure~\ref{fig:paretoerr:resnetgld_time}), PT has a lower median time penalty. PT does worse than NN only for \resnet with \mobilenet data (Figure~\ref{fig:paretoerr:mobnetin_time}).

\claim{PT optimization has the lowest normalized excess power AUC for most models and exceeds the power budget within a $1W$ threshold under $25\%$ of the time.}
In Figure~\ref{fig:pareto:barpower}, we report three metrics: Area, A/L and A/L + 1. Area stands for the \textit{normalized area under the curve (AUC)} of power in excess of the given budget for the given solution by a strategy (W per solution); A/L stands for the \textit{\% of solutions that exceeds the power limit} for a strategy; and A/L + 1 stands for the \textit{\% of solutions that exceed the limit by more than a 1-Watt threshold}.\\
As seen from Figure~\ref{fig:pareto:barpower}, the \textit{Area} is the lowest for PT across all DNN workloads except YOLO ($6$ out of $7$ cases). Additionally, this budget + 1W buffer (A/L + 1) is breached in $< 20\%$ of the solutions for $6$ out of $7$ DNNs and is $25\%$ for \mobilenet. So PT's predictions and solutions help stay within the power budget most of the time.

\claim{MAXN almost always exceeds the power budget while offering the best time, while Random sampling has higher time penalties with no power violations}
As seen from Figure~\ref{fig:pareto:boxtime}, MAXN has negative time penalties for \resnet, \mobilenet and \yolo, i.e., is much faster than the optimal solution. But this is because of setting all frequencies to the maximum value, thus violating the power limit constraints often (Figure~\ref{fig:pareto:barpower}).
In contrast, a Pareto from Random sampling is $12$--$28\%$ slower than the optimal. So these simpler baselines are sub-optimal and potentially unusable.

\section{Related Work}
\label{sec:related}

\subsection{Energy, Power and Performance Modeling on Edge devices}

Earlier works~\cite{matei_europar} use roofline modeling on a much older Jetson TK1 and TX1 to understand and characterize the performance of CPU and GPU micro-benchmarks for matrix multiplication. MIRAGE~\cite{mirage_sec} uses gradient tree boosting to predict runtime and power consumption for DNN inference on a Jetson Xavier AGX. Others~\cite{edge_config, igsc_edge} have investigated the impact of frequencies and cores with the latency, power and energy for inferencing on the Jetson Nano. Abdelhafez and Ripeanu~\cite{frqswitching} examine the effect of frequencies on power consumption for stream processing workloads. Some~\cite{iiwsc_tx1energy} have developed methods to measure fine-grained layer-wise energy for inference workloads on the Jetson TX1. All of these have either looked at inferencing or micro-benchmarks, and not DNN training.

Our previous work characterized training of DNNs on Jetson AGX, NX Xavier and Nano~\cite{sigmetrics23}. There, we offered initial insights on the opportunities for deep learning training on Jetson and some of the performance and power behavior. We also proposed a simple time and energy prediction model based on linear regression. But it is only evaluated on $\approx 10$ power modes, and its errors are much worse than what we observe in this work. 
In contrast, our goal is to model and predict the power and time for DNN training. 

\subsection{Energy, Power and Performance Modeling on GPU servers}
NeuralPower~\cite{NP_ACML} uses polynomial regression to predict power, runtime and energy consumption of CNNs for inference on server-grade GPUs \addc{as a function of the DNN architecture and their resource usage on a hardware. While it can generalize to other hardware, it needs to profile it initially. Our primary focus in this article is to predict training time and power for DNN training on Jetsons when these dimensions are modulated by the power modes. Each power mode effectively becomes a new hardware, causing each to be profiled by NeuralPower rather than predicted by us.}
Some~\cite{kernel_greencomp} use linear regression to predict the GPU power consumption of CUDA kernels based on hardware performance counters\addc{, but this cannot be applied to DNNs because of framework optimizations such as kernel fusion. Our prior work~\cite{sigmetrics23} shows that linear regression on these parameters is inadequate to make high quality predictions.} Paleo~\cite{paleo_ICLR} builds an analytical model of training time for DNNs based on the computational requirements of the DNN architecture, and maps them to the design space of software, hardware and communication strategies on server-grade GPUs. \addc{However, they too do not account for the hardware configurations such as frequencies, and are not applicable. }Others~\cite{powerperf_ipdps13} build a power, performance and energy model for application kernels on desktop or server-grade GPUs. Recent work~\cite{li_icdcs} builds regression models that predict the training speed in a distributed setup with cloud GPUs. All of these focus on server-grade hardware and they do not look into fine-grained power modes or frequencies along many dimensions, which we do. \addc{As a result, there are no other works that directly solve the problem that we address, and this also limits our choices for state-of-the-art baselines to emprirically evaluate PowerTrain against.}

\subsection{Optimization studies on the edge and server}
Zeus~\cite{zeus} studies the trade-off between training time and energy, and uses this to minimize the cost of a training job by selecting the best minibatch size and GPU power limit. Others~\cite{icpp23_gpufreq} select optimal GPU frequency that lowers power while minimizing the performance impact on server-grade GPUs. D3~\cite{stoica_d3} looks at optimizing the runtime accuracy trade-offs for DNN inference workloads in autonomous vehicle pipelines that have dynamically varying deadlines. ALERT~\cite{alert_atc20} selects a DNN and a power limit to meet accuracy, latency and energy constraints for inference tasks on laptop and desktop CPUs and GPUs. AxoNN~\cite{axonn_dac} distributes layers of a DNN inference workload between a performance-efficient GPU and a power-efficient DLA on the Jetson Xavier AGX to minimize time and stay within an energy budget. Others~\cite{kang2020scheduling} find the Pareto optimal scheduling of multiple deep learning applications in terms of the response time and energy consumption on smartphones. Mephesto~\cite{monil2020mephesto} models memory contention for kernel placement on heterogeneous SoCs to achieve the desired energy performance tradeoff. \addc{Oppertune~\cite{oppertune} addresses post-deployment optimization of applications by figuring out which parameters to tune and using reinforcement learning to tune both numerical and categorical variables. However, they often sample 100s of configurations, which is much costlier in our case and not usable. }We focus on modeling the training time and power for a DNN workload on an \textit{accelerated edge device} and solving an optimization problem using these models, which is a gap in the existing literature.

\section{Discussions and Conclusion}%\Note{0.5pgs}
\label{sec:conclusions}
In this work, we design two modeling techniques to predict the training time and power for DNN training workloads on Jetson edge devices. PowerTrain uses hours of offline data collection and rigorous training for a reference DNN workload to bootstrap the prediction model for a whole new DNN workload within a few minutes of profiling for $\approx 50$ power modes and retraining. PowerTrain's power and time predictions generalize well across overlapping DNNs and datasets, unseen DNN workloads and even new devices. We leverage PowerTrain to come up with a predicted Pareto to trade-off power limits against training time for a given DNN workload training. This helps identify the optimal power mode to stay within a given power limit while minimizing the training time for the DNN workload. Our results confirm the robustness of PowerTrain to be reused in diverse workload and device conditions with low overheads, and its out-performance on predictions and in solving the optimization project compared to simple and NN baselines. This makes it a valuable methodology for efficient configuration of dynamic DNN training workloads in edge deployments.

In future, PowerTrain can be used in an online fashion for practical deployment, allowing us to collect sampling data for ML training while the DNN workload runs productively. We also plan to explore reinforcement learning based methods to solve this problem. We plan to explore variants of this problem, such as time and power prediction for concurrent training and inference workloads \addc{and power-aware minibatch size optimization for training workloads}. This work can also be investigated for use in \delc{other edge devices like Rapberry Pi and }GPU servers, and also non-DNN workloads.

\section*{Acknowledgments}
The authors would like to thank students in the DREAM:Lab including Pranjal Naman, Suman Raj and Kedar Dhule for their assistance with the article. The first author was supported by a Prime Minister's Research Fellowship (PMRF) from the Ministry of Education, India.

\balance

\bibliographystyle{IEEEtran}
\bibliography{arxiv}

\clearpage

\appendix

\addc{\section{Appendix}}

\begin{table*}[t]
\footnotesize
\caption{\addc{Specifications of Server GPUs and Edge Devices}}
\label{tbl:hw_var_specs}
\begin{minipage}[t]{.8\columnwidth}
\centering
\begin{tabular}{L{2.3 cm}|L{2 cm}|L{1.7 cm}|L{1.8 cm}|L{2 cm}}
\hline
\textbf{Feature} & \textbf{3090} & \textbf{A5000} & \textbf{Orin AGX} & \textbf{Raspberry Pi5} \\
\hline\hline
CPU  & 12th Gen Intel i9-12900K & AMD EPYC 7532 & ARM Cortex A78AE & ARM Cortex-A76 \\ 
\hline
\# CPU Cores & 16 & 32 & 12 & 4 \\
\hline
Max CPU frequency (MHz) & 5200 & 3400 & 2200 & 2400 \\
\hline
GPU (arch/CUDA cores) & Ampere, 10496 & Ampere, 8192 & Ampere, 2048 & VideoCore VII (graphics only) \\
\hline
Max GPU frequency (MHz) & 1695 & 2505 & 1300 & 800 \\
\hline
RAM (GB) & 128 (CPU) + 24 (GPU) & 512 (CPU) + 24 (GPU) & 64 (shared) & 8 \\
\hline
Peak Power (W) & 350 & 230 & 60 & 27 \\
\hline
\end{tabular}
\end{minipage}
\end{table*}

\begin{figure*}[b]
\centering
\includegraphics[width=1\textwidth]{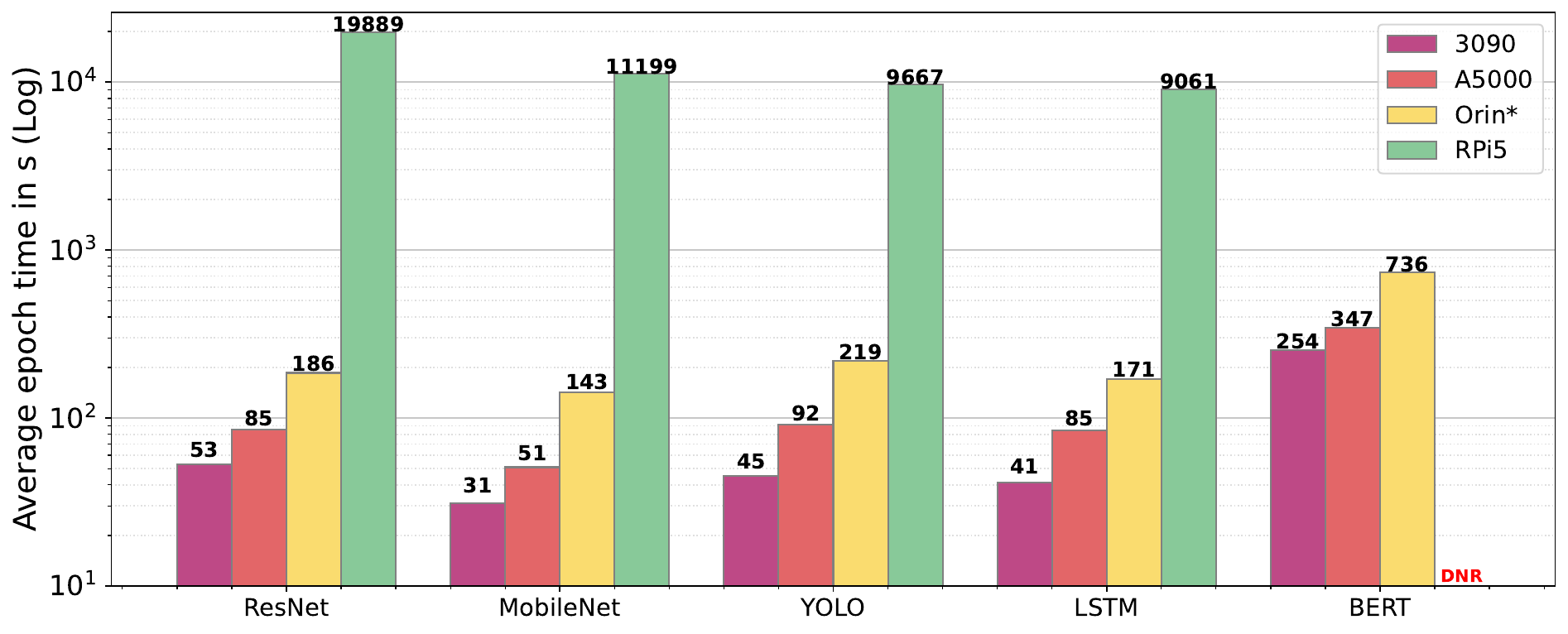}
\caption{\addc{Average epoch times across various edge and server devices. The device specifications are: \textbf{3090} \textit{(12th Gen Intel i9 CPU with 16 cores@ 5.2GHz, 128GB RAM, 3090 Ampere GPU with 10496 cores, 24GB GPU RAM)}; \textbf{A5000} \textit{(AMD EPYC 753 CPU with 32 core@ 3.4GHz, 512GB RAM, A5000 Ampere GPU with 8192 cores, 24GB GPU RAM)}; \textbf {Orin} \textit{(ARM A78AE CPU, 12 cores@ 2.2GHz, Ampere GPU with 2048 cores, 32 GB shared RAM)}; \textbf{RPi5} \textit{(ARM A76 CPU, 4 cores@ 2.4GHz, 8GB RAM)}}}
\label{hw_var_train}
\end{figure*}

\addc{As a point of reference to compare the compute power of these Nvidia Jetson devices, we ran the $5$ training workloads on $3$ additional device types and contrast their compute performance against the Nvidia Jetson Orin device used in our experiments~\footnote{We use the same library versions for these new devices (LTS versions; PyTorch 2.3, Ultralytics 8.2.25, CUDA 12.1). 
 % for the 3090 and A5000 \ysnote{
But these are newer than the versions in Orin AGX due to Nvidia JetPack's slower update cycle.}: (1) A GPU server with NVIDIA RTX A5000~\cite{A5000}, (2) A GPU workstation with NVIDIA RTX 3090~\cite{RTX3090}, and (3) The latest generation (non-accelerated) Raspberry Pi5 edge device~\cite{rpi5} and report the average epoch times in Figure~\ref{hw_var_train}. The specifications of the devices are in Table~\ref{tbl:hw_var_specs}.
% update frequcilimitations. 
We observe that the 3090 is significantly faster than the A5000, which can be explained by the 3090 having more CUDA cores of the same Ampere generation ($10,496$ vs. $8,192$). The Orin AGX is slower than the A5000 and the 3090, as expected, due to fewer Ampere CUDA cores (2048). The Raspberry Pi 5 trains using just the ARM CPU cores, and is two orders of magnitude slower than the Orin. BERT was unable to run on the RPi5 (\texttt{DNR} in Figure~\ref{hw_var_train}) as it ran out of memory in 8GB of memory available in the RPi.}

\end{document}